\documentclass[12pt]{article}
\usepackage{epsfig}
\usepackage{amstex}
\usepackage{amssymb}
\def\lambdabar{\protect\@lambdabar}
\def\@lambdabar{
\relax
\bgroup
\def\@tempa{\hbox{\raise.73\ht0
\hbox to0pt{\kern.25\wd0\vrule width.5\wd0
height.1pt depth.1pt\hss}\box0}}
\mathchoice{\setbox0\hbox{$\displaystyle\lambda$}\@tempa}
{\setbox0\hbox{$\textstyle\lambda$}\@tempa}
{\setbox0\hbox{$\scriptstyle\lambda$}\@tempa}
{\setbox0\hbox{$scriptscriptstyle\lambda$}\@tempa}
\egroup}
\begin{document}
\title{Lasers as a Bridge between Atomic and Nuclear Physics}
\author{Sergei Matinyan\\Duke University, Physics Department\\and\\Yerevan 
Physics Institute, Armenia}
\maketitle
\section*{Abstract}
This paper reviews the application of optical and UV laser radiation to 
several topics in low-energy nuclear physics.  We consider the 
laser-induced nuclear anti-Stokes transitions, the laser-assisted and the 
laser-induced internal conversion, and the Electron Bridge and Inverse 
Electron Bridge mechanisms as tools for deexciting and exciting of low-lying 
nuclear isomeric states. A study of the anomalous, by low-lying, nuclear 
isomeric states (on an example of the $^{229}$Th nucleus) is presented in 
detail.\\
PACS: 21.10.-k, 23.20.Lv, 23.20.Nx, 32.80.-t, 32.80.Wr, 42.62.-b\\

\vspace{5mm}

Keywords:  laser-assisted and laser-induced nuclear transitions; Electron 
Bridge and Inverse Electron Bridge mechanisms; excitation and deexcitation of
nuclear levels; nuclear isomers; anomalously low-lying nuclear levels.

\newpage
\tableofcontents
\newpage

\section*{Preface}
\addcontentsline{toc}{section}{\numberline{}Preface}

The present paper reviews the direct application of laser radiation to 
several topics in nuclear physics.

Here the term ``direct'' refers to the use of laser radiation in the optical 
and UV region (energy of the laser photons 
$\hbar \omega_L \sim$ 1-10 eV ($\lambda_L = 10^{-4} - 10^{-5}$ cm)) for  
studying low-energy nuclear phenomena and related processes.

The ``traditional'' approach of using laser radiation in  
nuclear physics is well known. The Compton backscattering of 
laser photons off accelerated electrons produces intense, highly ($\sim$ 100\%)
polarized $\gamma$-ray beams with excellent energy resolution. This very 
effective method of employing conventional lasers in particle and nuclear 
physics was proposed long ago \cite{FRA} and was implemented successfully in 
various accelerator facilities (SLAC, 
BNL, LURE (Orsay), ETL (Tsukuba)).  The recent completion of the Free Electron
Laser (FEL) Facility at Duke University gives possibility to yield 
$\gamma$-ray fluxes of $\approx 10^7-10^9 {\gamma \over s}$ (a factor of $10^3$
higher than is obtainable with conventional lasers in the energy range from 
2-200 MeV \cite{TCA} for studying phenomena 
of low and intermediate energy nuclear and particle physics.

In the present survey these kinds of topics in Nuclear Physics will be 
completely avoided. The aim here is to consider the wide range of very 
low-energy nuclear 
physics topics where the ``direct'' (i.e., not transformed into $\gamma$-rays 
by Compton backscattering) laser 
beam\footnote{For the qualitative features of the interaction
of the laser radiation with electron see Appendix A.} interaction with 
electronic atomic shells serves as a tool for studying the properties of 
low-lying nuclear levels and accompanying dynamical processes.

The theoretical study of these phenomena has a long and sometimes controversial
history.  It is interesting to point out that the history of these studies 
dates
back to the 1920's when Einstein discussed the possibility that radioactivity 
could be induced by the action of \underline{optical} quanta \cite{AEI}.  

The literature on the subject of the present review is comprised of more than 
one hundred studies published mostly in American and former Soviet physical
journals. To my knowledge there has not been an attempt to review 
this field exhaustively.  Unfortunately, in the review literature, the 
main papers from the former USSR, which contribute significantly to the field,
are not discussed appropriately. The short review paper \cite{PKA} suffers 
from one-sidedness.

Motivated by these reasons and stimulated by some of my colleagues from 
Triangle Universities Nuclear Laboratory (TUNL) and
the Physics Department of Duke University, I present a rather
complete review of the field covering the ``direct'' use of 
laser radiation in the energy range of the FEL Facility at Duke University 
($E_{L} \lesssim 12$ eV) for studying low-energy nuclear physics phenomena.

Due to the diversity of the topics and methods, the present survey has to be 
rather schematic, but this will be compensated for by providing a 
detailed list of references.

The paper is organized as follows:

After introductory remarks concerning the ineffectiveness of laser
radiation in the optical range for the unmediated study of properties
of nuclei, the role of the atomic shell as a mediator is shown in the case
of Compton excitation of nuclei.  In Section 3, the laser-induced nuclear 
anti-Stokes transitions are investigated and electromagnetic transitions are
compared with $\beta$-transitions.  It is shown that the laser-induced 
electromagnetic anti-Stokes transitions are effective only for very strong 
laser fields ($I \gtrsim 10^{19} {W \over cm^2}$).  Section 4 is devoted to 
the laser-assisted internal conversion process.  The laser plays an important 
role in specific nuclei where the internal conversion in the laser-free case 
is forbidden by energy conservation.  The technique for calculating the 
internal conversion coefficient for this process is presented.  Experimental 
issues related to this process are discussed.
In Section 5 we study the Electron Bridge mechanism as a means for 
deexciting nuclei and we discuss the role of laser radiation in enhancing 
the deexcitation.  The Inverse Electron Bridge mechanism is presented as a 
tool for exciting nuclear states to their isomeric states. 
Section 6 is devoted to the study of the anomalously low-lying nuclear isomeric
states in the $^{229m}Th$ nucleus.
Methods introduced in Sections 4 and 5 are used for the detailed study of this
phenomenon.
Finally, Section 7 gives a summary.

In Appendix A dimensional arguments are used to give some useful information 
about the role that the high intensity plays in the interaction of radiation
with electrons and atoms.
Finally, Appendix B deals with the important issue of two alternative ways of 
describing the interaction of radiation with matter.
\newpage

\section{Introduction}
\hfill
\parbox{2in}{\footnotesize Q.  Is it possible to obtain induced radioactivity by bombardment of matter
with quanta of light?\\
A. First of all, I have to say that, probably, there exists radioactivity of
matter induced by the action of the light quanta; the difficulty of the 
observation of such phenomenon, if it exists, is that the effect which has to
be observed is very small.
The confirmation of this effect is hard but possible.\footnotemark}
\footnotetext{On April 16, 1925, the National Academy of Science of Argentina
 held a special session to award A. Einstein the Academy's honorary member 
diploma.

At the reception, Einstein answered the questions of the Academy members.  Four
years later, details of the Session and the ``interview'' were published in 
the Proceedings of the Academy (Reception de la doctor Alberto Einstein en la 
session especial de la Akademia el dia 16 de april de 1925; Annales Sociedad 
Cientifica Argentina, \underline{107}, 337-347 (1929)) \cite{AEI}.}

\subsection{Simple estimate}

Since the wavelength of the laser radiation $\lambda_L$ is much larger than 
the 
nuclear size $R_A$ [$\sim (2-10) \times 10^{-13}$ cm], the direct coupling 
strength
of laser photons to a nuclear system considered as a quantum state with level
spacing $\Delta E_A$ of the order of one hundred keV, is very small, even at 
extremely high radiation intensity.  It is worthwhile, to appreciate this 
smallness, to make the following simple estimate.
The interaction of electromagnetic radiation with a nucleus is more 
conveniently described by the Hamiltonian\footnote{The problem of relating the
 dipole from (1) to
the commonly used form of the matter-field interaction Hamiltonian
${1 \over 2m}(\vec{p} - e\vec{A})^2 \rightarrow -{e \over m}\vec{p} \cdot
\vec{A} + {e^2 \over 2m}\vec{A}^2$, has a long and controversial history which
is resolved in the paper of Ref. \cite{WEL}. See Appendix C for some details.}
in the so-called G{\"o}ppert-Mayer's gauge\cite{MGO}

\begin{equation}
H_{LA} = -e \sum_{p=1}^Z  \vec{r_p} \cdot \vec{E_L}(t), 
\end{equation}

\noindent
where $\vec{r_p}$ is the radius vector of individual protons in the nucleus 
and $\vec{E_L}$ 
is
the electric field of the laser radiation.  $|\vec{E_L}|$ is proportional
to $I^{1 \over 2}$ where I is the laser beam intensity which will be 
measured hereafter in units of ${W \over cm^2}$.

>From (1) the following estimate for the magnitude of the matrix element
can be derived:

\begin{equation}
{\mathbf M} \approx (10^{-12} - 10^{-13})I^{1 \over 2} \; \text{eV}.
\end{equation}

\noindent
Using for the nuclear level spacing $\Delta E_A \gtrsim$ 1 keV, we obtain for 
the
coupling strength in the best case scenario (e.g., when the shielding of the 
atomic
shells is ignored)

\begin{equation}
{{\mathbf M} \over \Delta E_A} \lesssim (10^{-15} - 10^{-16})I^{1 \over 2}.
\end{equation}

The situation for the direct use of laser radiation remains hopeless even
in the case of very low lying levels (e.g., for $^{229}Th$, where $\Delta E_A$
is on the order of a few eV.  See Section 6).

\subsection{Nuclear Compton Effect}
Here we will show that the direct excitation of nuclear levels by X-rays or by
hypothetical $\gamma$-ray lasers does not look encouraging either.

Consider, for example, the nuclear excitation by the Compton effect at 
resonance (See Fig. 1 for the corresponding Feynman diagrams).

For the cross section we have in the one resonance level approximation:

\begin{equation}
\sigma_{\gamma\gamma}^A = {C_N \over k_L^2}{\Gamma_{IN}\Gamma_{NF} \over 
(E_{IN} - \hbar\omega_L)^2 + {\Gamma^2 \over 4}} \text{,}
\end{equation}

\noindent
where $\Gamma_{IN}$ and $\Gamma_{NF}$ are the widths of level $N$ associated
with the transition to the initial state, $I$, and to the final state, $F$,
respectively.
Here, $k_L(\omega_L)$ is the wave number (frequency) of the incident laser 
photon, 
$C_N$ is the statistical weight of the intermediate state $N$, and $\Gamma$ is
the total width of the level $N$.  $E_{IN} \equiv E_I - E_N$, where $E_{I(N)}$
is the energy of the state $I(N)$.

Estimates \cite{ISB} with $E_F - E_I - \hbar\omega_L \approx 10^5 eV$
give for the cross section
$$\sigma_{\gamma\gamma}^A \approx 10^{-39} - 10^{-40} cm^2,$$
which practically is unobservably small.

These introductory examples show the ineffectiveness of direct excitation of 
nuclear levels by radiation.

\section{Atomic Shells in the ``Compton'' Excitation of Nuclei}
The example considered in the Introduction suggests the following modification
to the Compton mechanism involving the excitation of low-lying nuclear levels:
when the Compton effect occurs on the electron bound in the atomic shell,  
excitation of the nucleus takes place instead of scattering of the photon.

The mechanism is described by the diagrams in Figs. 2a and 2b.  We note 
that if one 
exchanges the direction of the photon line (absorption of a photon is 
changed to emission of a photon in the final state) one arrives at the 
so-called
internal Compton effect where the photon from the deexcitation of the nucleus 
is scattered on a bound electron.

This last mechanism is the basis of the Electron Bridge (EB) mechanism whereas
the first one (described by the diagrams in Fig. 2) corresponds to the Inverse
Electron Bridge (IEB) mechanism. 
The EB mechanism can be considered as alternative to the well-known 
Internal Conversion (IC) process where the deexcitation of the nucleus results
in ejection of an electron out of the atomic shell without emission of a 
photon.

We will see below how laser radiation can assist in influencing the EB and IC 
mechanisms by resonating at the corresponding atomic level or by eliminating 
the mismatch between the nuclear ($\Delta E$) and atomic ($\Delta\epsilon$) 
level differences.

Here, two aspects of laser radiation are very important:\\
i) The dipole character of the interaction with the electronic shell which, of
course, is common to all kinds of electromagnetic radiation.  This property 
leads to an effective 
reduction of possible multipolarities inherent, in general, to nuclear 
radiation;\\
ii) unique, due to the lasers' very high intensity, is the possibility of 
multi-photon absorption (emission). This fact facilitates the resonance 
matching between $\Delta E$ and $\Delta\epsilon$. 

These two essential aspects of laser radiation will be focused on below
in the study of the role of laser radiation in low energy nuclear physics.

Turning now to the excitation of low-lying nuclear states
(generally nonresonant) by $\gamma$-rays (diagrams in Figs. 2a and 2b),
we write the matrix elements $\mathbf{M_a}$ and $\mathbf{M_b}$ 
corresponding to Figs. 2a and 2b:

\begin{equation}
{\mathbf M}_a = \sum_n {\langle fF|H_{int}|In\rangle \langle 
n|H_{\gamma}|i\rangle
\over \epsilon_n - \epsilon_i -\hbar\omega_L} \tag{5a}
\end{equation}

\begin{equation}
{\mathbf M}_b = \sum_n {\langle f|H_{\gamma}|n\rangle \langle 
nF|H_{int}|iI\rangle 
\over \epsilon_n - \epsilon_f + \hbar\omega_L} \tag{5b}
\end{equation}

We have used the following notations: $|i\rangle$, $|n\rangle$, $|f\rangle$  
describe the initial, intermediate, and final states of an electron with 
energies $\epsilon_i$, 
$\epsilon_n$, $\epsilon_f$  respectively. $I$ and $F$ denote the 
initial and final states of the nucleus with energies $E_I$ and $E_F$. 
$H_{int}$ 
is the Hamiltonian describing the interaction of an electron with a nucleus 
and $H_{\gamma}$ is 
the Hamiltonian describing the interaction of a photon with an electron.\\
Conservation of energy gives: $\epsilon_i + \hbar\omega_L = \epsilon_f + 
E_{FI}$
($E_{FI} \equiv E_F - E_I$).\\
We confine ourselves here to the non-covariant perturbation theory (see(5a,b))
since the exact calculation requires construction of the relativistic Green's 
function of an electron in the field of an atom.  The physically simpler 
treatment above is transparent and effective.

We concentrate on the case where the laser photon energy is much smaller
than the rest energy of the electron, $\hbar\omega_L < mc^2$ or, more 
precisely, smaller than the electron binding energy in the shell, 
$\hbar\omega_L \lesssim
(\alpha Z)^2mc^2$, where $\alpha = {e^2 \over \hbar c}$ is the fine structure
constant, m is the electron mass, and Z is the atomic number.

Inspection of the denominators in the matrix elements (5a,b) shows
that the main contribution to $\mathbf{M_a}$ results from the summation over 
positive energy $\epsilon_n$
values of the states $|n\rangle$, whereas the 
main contribution
to $\mathbf{M_b}$ comes from negative values of $\epsilon_n$.  That means 
that
the excitation of the nucleus described by diagram 5a occurs mainly through 
inelastic scattering of a free electron ($\epsilon_n > 0$) (free-free 
transitions).
Therefore, diagram 5a gives the main contribution at sufficiently high
incident photon energies ($\hbar\omega_L \gg (\alpha Z)^2mc^2$).  On the other
hand,
diagram 5b is more important when the nucleus is excited in transitions 
of the electron between bound-bound and bound-free atomic states.  This 
indicates
that this diagram gives the main contribution at the energies we are 
interested
in ($\hbar\omega_L \lesssim (\alpha Z)^2mc^2$).  Calculations \cite{ISB} 
indeed confirm these physical reasonings. 

In the calculation of the cross section
\addtocounter{equation}{1}
\begin{equation}
\sigma_{\gamma} = {2\pi \over \hbar}{1 \over 2J_I + 1} \sum_i {1 \over 2J_i + 
1}
\sum_{m_im_f} \sum_{m_Im_F} |M_b|^2\rho(\epsilon_f)
\end{equation}
(here $\rho(\epsilon_f)$ is the density of the states of the final electron, 
$m_i$, $m_f$ and $m_I$, $m_F$ are magnetic quantum numbers of the initial and
final states of the electron and the nucleus, respectively; the summation 
index 
$i$ denotes the occupied states in the (initial) electron shell) the
approximation of one resonance level n \cite{ISB} is used which is more 
effective and 
reasonable in the case of bound-bound transitions.  If the energies of the
nuclear excitation and of the electronic transition are sufficiently close,
a significant increase of the cross section is achievable.

The resulting cross section includes the probability of $\gamma$-radiation of 
multipolarity $L$ emitted by the nucleus, and the well-known matrix elements 
for the electronic electromagnetic transitions $\mathbf{M}(EL)$ 
and $\mathbf{M}(ML)$ ($E$ and $M$ in $\mathbf{M}$ refer to electric and 
magnetic transitions, respectively). 
For higher multipolarities, the cross section drops sharply 
($\sim ({R_A \over a})^{2L}$, where $R_A$ is the radius of the nucleus and $a$ 
is the 
radius of the electron shell).  Thus, the maximum value of $\sigma_{\gamma}$
occurs for transitions with the lowest possible multipolarity $L$.  Expressions
for the matrix elements
$\mathbf{M}(E(M), L)$ are given in the literature (see, e.g.\cite{IMB}).  The 
summation in 
(6) is carried out over the occupied states in the initial and final electron 
shells, and the choice of 
the characteristics of the intermediate state $|n\rangle$ (e.g., its
angular momentum $j_n$, etc.) depends on the initial state $|i\rangle$.  In 
the calculations the widths of the atomic and nuclear levels are usually 
neglected in comparison with 
the parameter $\Delta \equiv E_I - E_F + \epsilon_n - \epsilon_i$. This 
parameter 
enters in the cross section in the form $[{(E_I - E_F) \over \Delta}]^2$.
>From this, it is evident that for the bound-free transition case, 
$\epsilon_i < \epsilon_n$, $\Delta \approx E_I - E_F$, i.e., the resonance 
condition is not fulfilled.

The largest cross section occurs for the bound-bound atomic transitions.  This
case is most effective if $E_{IF}$ is sufficiently close to the energy of 
the 
electronic transition ($\Delta$ is small).  $M1$ transitions are very 
important.  For example, the cross section of the Compton excitation of the
isotype $_{50}^{110}Sn$ with the $M1$ transition ${1 \over 2}^+ \rightarrow 
{3 \over 2}^+$, $E_{FI} = {24}$ keV, the atomic transition $2s_{1 \over 2} 
\rightarrow 
1s_{1 \over 2}$ and $\Delta$ = 0.74 keV, one obtains $\sigma_{\gamma} = 
4\mu b$.  In
Table I, we show the parameters for nuclear and atomic transitions and the
resulting cross sections $\sigma_{\gamma}$ for several long-lived isotopes.
The energies of the nuclear transitions have been taken from \cite{BSD} and
those of the electronic transitions from \cite{JAB}.  We see that the 
magnitudes of the cross section are considerable ($\sim (10 - 10^2)\mu b$).

The case of the bound-free transitions is less effective. Here,  
$\sigma_{\gamma}$ is
of the order of($10^{-2} - 1$)$\mu b$.  For example \cite{ISB}:\\
$_{26}^{57}Fe$, $\sigma_{\gamma} = 5 \cdot 10^{-2}\mu b$; \quad $_{36}^{83}Kr$,
$\sigma_{\gamma} = 10^{-1}\mu b$; \quad $_{54}^{129}Xe$, $\sigma_{\gamma} = 
1 \mu b$.\\
The conclusion from the above considerations is that the relatively low lying
nuclear vibrational, collective states can be excited at a rate of($10^{-7} - 
10^{-9}$) 
per photon.

\vspace{5mm}

\begin{center}
\begin{tabular}{|c|c|c|c|c|c|c|}\hline
Isotope & $E_{FI}$, keV & $J_I \rightarrow J_F$ & $j_i \rightarrow j_n$ & 
$\epsilon_{j_nj_i}$, keV & $\Delta$, keV & $\sigma_{\gamma}$, $\mu b$\\
\hline
$_{50}^{119}Sn$ & 24 & ${1 \over 2}^+ \rightarrow {3 \over 2}^+$ & 
$2s_{1 \over 2} \rightarrow 1s_{1 \over 2}$ & 24.74 & 0.74 & 4\\
\hline
$_{53}^{129}J$ & 27 & ${7 \over 2}^+ \rightarrow {5 \over 2}^+$ &
$2s_{1 \over 2} \rightarrow 1s_{1 \over 2}$ & 27.88 & 0.88 & 5\\
\hline
$_{52}^{125}Te$ & 36.5 & ${1 \over 2}^+ \rightarrow {3 \over 2}^+$ &
$3s_{1 \over 2} \rightarrow 1s_{1 \over 2}$ & 30.81 & -4.7 & 0.14\\
\hline
$_{66}^{161}Dy$ & 43.8 & ${1 \over 2}^+ \rightarrow {3 \over 2}^+$ &
$2s_{1 \over 2} \rightarrow 1s_{1 \over 2}$ & 44.74 & 0.44 & 110\\
\hline
$_{76}^{185}Tm$ & 8.42 & ${1 \over 2}^+ \rightarrow {3 \over 2}^+$ &
$3s_{1 \over 2} \rightarrow 2s_{1 \over 2}$ & 7.81 & -0.61 & 196\\
\hline
$_{76}^{189}Os$ & 69.60 & ${3 \over 2}^- \rightarrow {5 \over 2}^-$ &
$3s_{1 \over 2} \rightarrow 1s_{1 \over 2}$ & 70.82 & 1.22 & 26\\
\hline
$_{77}^{193}Ir$ & 73.1 & ${3 \over 2}^+ \rightarrow {1 \over 2}^+$ &
$3s_{1 \over 2} \rightarrow 1s_{1 \over 2}$ &  72.85 & -0.25 & 9\\
\hline
$_{79}^{197}Au$ & 77 & ${3 \over 2}^+ \rightarrow {1 \over 2}^+$ &
$3s_{1 \over 2} \rightarrow 1s_{1 \over 2}$ & 77.4 & 0.4 & 10\\
\hline
$_{80}^{201}Mg$ & 1.57 & ${3 \over 2}^- \rightarrow {5 \over 2}^-$ &
$2p_{3 \over 2} \rightarrow 2p_{1 \over 2}$ & 1.93 & 0.36 & 17\\
\hline
$_{78}^{193}Pt$ & 12.7 & ${1 \over 2}^- \rightarrow {3 \over 2}^-$ &
$4s_{1 \over 2} \rightarrow 2s_{1 \over 2}$ & 13.04 & 0.34 & 62\\\hline
\end{tabular}
\end{center}

\vspace{5mm}

Table I. Parameters of nuclear ($E_{FI}$ (keV), $J_I \rightarrow J_F$)
and atomic ($\epsilon_{j_n,j_i}$, $j_i \rightarrow j_n$) bound-bound
transitions, $\Delta = E_I - E_F + \epsilon_j - \epsilon_{j_i} = -E_{FI} + 
\epsilon_{j_nj_i}$, and calculated Compton excitation cross sections
for some long-lived isotopes \cite{ISB}.

\vspace{5mm}

Of course, the experimental situation is very difficult, due to the high 
background
associated with the intense primary and secondary fluxes of X-rays. 
The fact that the energies of the X-rays and the nuclear radiations do not 
coincide (the difference reaches several keV) can perhaps be used to 
detect and to study low-lying states in nuclei.  Concluding, we mention for the
sake of completeness, that in the case of higher energy  
$\gamma$-rays ($\hbar\omega_L \gg (\alpha Z)^2mc^2$), where diagram 5a
gives the main contribution, the cross sections are much lower, typically of 
the order of $10^{-34}cm^2$.

The above examples which do not always have a direct connection with the 
nuclear
physics conducted by UV and optical lasers (the main topic of interest in the
present work),  teaches
us that atomic physics is merged with nuclear physics due to the coupling
of electromagnetic radiation with the electronic shells.  Lasers with their
high intensity have the potential to enhance and sometimes induce
nuclear (weak and electromagnetic) transitions.

Below we will illustrate these issues in several examples (Sections 3-6).

\section{Laser induced nuclear anti-Stokes transitions}

\subsection{$\beta$-transitions of nuclei}
Lasers, like other sources of electromagnetic radiation, provide the 
possibility
of enhancing transitions from excited nuclear states and of obtaining 
information on specific nuclear transitions unaccessible otherwise.

Consider, as an example, a three-level nuclear system as illustrated in
Fig. 3.  First, we concentrate on the $\beta$-transition:  The
nucleus initially is in the isomeric state $a$ from which it 
$\beta$-decays (very slowly) to its ground state $b$ with the
rate $\gamma_{ab}^w$.  Under the influence of an external electromagnetic 
field with frequency $\omega_L$ one can induce a two-step $\beta$-decay of 
$a$ to $b$: a virtual excitation of the nucleus to the
level $c$ by absorption of a single (or, sometimes multiple) photon
with energy $\hbar\omega_L$ and rate $\gamma_{ca}^{el}$ of the laser field
and subsequent $\beta$-decay from $c \rightarrow b$
with rate $\gamma_{cb}^w$.

This scheme gives not only the possibility of increasing the rate of the
$\beta$-decay to the state $b$ (if $\gamma_{cb}^w > 
\gamma_{ab}^w$; of course, $\gamma_{cb}^w \ll 
\gamma_{ca}^{el}$) but provides the information on the $\beta$-transition
$c \rightarrow b$ which is sometimes not accessible \cite{WBE}.

Before turning to the more interesting cases where all the transitions 
discussed above are of 
electromagnetic nature, thus eliminating the condition $\gamma_{cb}^w \ll 
\gamma_{ca}^{el}$ (unavoidable for the $\beta$-transitions) it is 
worth mentioning that claims (1977-1983) about strong modifications of 
forbidden and even allowed $\beta$-decay \underline{total} rates \cite{MRR} 
by very intense,
long-wavelength, coherent electromagnetic radiation were
not confirmed, although their discussion has attracted the interest of many
authors.
The latest investigations (1987) have demonstrated that there is no laser
field-dependent enhancement of the total $\beta$-decay rates in the 
long-wavelength limit, although the spectra are greatly modified (for a
more complete review and the resolution of the controversies
in some previous studies see \cite{JLF,IMT}). However, we note that, in 
principle, the 
rate of another nuclear $\beta$-process, the orbital electron capture by a
nucleus (so called K-capture) is naturally influenced by the laser radiation,
because the probability of this process depends on the electron density in the
vicinity of the nucleus. Therefore, the K-capture process can be sensitive to 
the influence of a strong electromangetic field.  
Particularly, in contrast to the case of the $\beta$-decay
where the total rate does not depend on the nuclear spin 
orientation\footnote{Of course, due to the parity non-conservation, the
probability of the particle ($e^-$ or $\nu_e$) emission parallel
to the nuclear spin in $\beta$ decay differs from the probability for the 
emission in the opposite direction.  However, the \underline {total} rate does
not depend on nuclear spin orientation simply because of the isotropy of 
space (rotational invariance).}, for the orbital electron capture, due to 
the hyperfine interaction between the electron and the nucleus, the total
rate of K-capture depends critically on the orientation of the electron
and nuclear spins, and therefore, on temperature \cite{LMF}.

The rate of the orbital electron capture from 
an unfilled $s$ orbital of a free atom or ion at low temperature ($kT \ll 
\delta$, 
where $\delta$ is the hyperfine splitting) depends on the sign of the 
hyperfine constant $h$ \cite{LMF}.
The application of a laser field with the frequency $\omega_L = 
{\delta \over \hbar}$ can influence the rate.  In the 
particular case of negative $h$, when the orbital electron capture practically 
does not take place, 
it is possible to induce this  process by applying the resonant radiation with
$\omega_L = {\delta \over \hbar}$.  
Large effects are expected especially for hydrogen-like ions since the 
competing decays from other bound electrons are absent.\footnote{The orbital 
electron capture probability rapidly decreases with an increase in the number 
of the competing atomic shells from which an electron can be captured.}
In principle, properly chosen radiation could cause a modulation of the 
electron capture rate and the subsequent characteristic X-ray emission could 
serve as a monitor for observing this process.

As a justification of the old wisdom that 
``the novel is well forgotten olden'' and
for the sake of completeness, we note that a very similar idea was proposed
almost two decades ago \cite{ISB2} for modifiying the population of hyperfine
structure sub-levels of $\mu$-mesic atoms through resonant laser radiation, 
and the subsequent change in the nuclear $\mu$-capture probability. 

\subsection{Electromagnetic transitions of nuclei}
Turning now to the case of electromagnetic nuclear transitions, we 
again
consider the three-level nuclear system with electromagnetic transition rates 
$\gamma_{ab}$, $\gamma_{ca}$, and $\gamma_{cb}$ and an applied laser field
of frequency $\omega_L$ to promote the $a \rightarrow c$ transition (Fig. 4) 
from the isomeric state $^ma$ to the level $c$.

It is easy to recognize that this scheme resembles the scheme of the famous
Lamb-Retherford experiment on the Lamb shift measurement of the hydrogen atom
where the level $a$ corresponds to the $2s_{1 \over 2}$ atomic level, the 
level 
$b$ corresponds to the level of $1s_{1 \over 2}$ ground state, and level
$c$ corresponds to the $2p_{1 \over 2}$ state which is excited by electron 
beam bombardment.

The fact that, in contrast to the $\beta$-decay, $\gamma_{ca} \approx 
\gamma_{cb}$, opens new possibilities.
The state $c$ can decay to states $a$ and $b$ by
the emission of radiation and also by internal conversion (IC)
to other, not specified states with total decay rate
$\gamma_c = \gamma_{ca} + \gamma_{cb} + \gamma_{IC} + \ldots$.

We take the interaction between the external radiation field and the system 
from (1) with $\vec{E_L} = \vec{E_0}\cos \omega_Lt$.  The laser field 
can be considered
homogeneous (spatially uniform) as its wavelength is much larger than 
atomic (and, of course, nuclear)
dimensions.

The equations of motion for the probability amplitudes of states $a$ and 
$c$ in the interaction representation (so called, Bethe-Lamb equations) 
\cite{MAB,Sar74,Bec84,WEL} are :

\begin{eqnarray}
\dot{a} = -{1 \over 2}\gamma_{ab}a - {i \over 2\hbar}V_{ca}e^{-i\Delta t}
\notag \\ 
 \\
\dot{c} = -{1 \over 2}\gamma_cc - {i \over 2\hbar}V_{ac}e^{i\Delta t} \notag
\end{eqnarray}

\noindent
where $V_{ac}$ is the transition dipole matrix element between the states $a$
and $c$, and $\Delta = {E_c - E_a \over \hbar} - \omega_L$ is a detuning
frequency.
In the following, we concentrate on the most interesting case where the 
spontaneous decay
of state $c$ to the isomeric state $a$ is negligible 
($\gamma_{ca} \approx 0$, no backcoupling $c \rightarrow a$).

Solving then Eqs. (7) with the initial conditions $a(0) = 1$, $c(0) = 0$, we 
have:
$$|a(t)|^2 = e^{-\gamma_{ab}t}$$

\begin{equation}
|c(t)|^2 = ({|V_{ac}| \over 2\hbar})^2 {e^{-\gamma_{ab}t} + e^{-\gamma_ct} 
- 2e^{-\gamma t}
\cos \Delta t \over \Delta^2 + \delta^2}
\end{equation}

\noindent
with

\begin{equation}
\begin{pmatrix}\gamma \\ \delta\end{pmatrix} = {(\gamma_{ab} \pm \gamma_c) 
\over 2}
\end{equation}

Since we are interested in the case where the state $c$ decays much faster 
than the 
isomeric state $a$ ($\gamma_c \gg \gamma_{ab}$), $\gamma = -\delta = 
{\gamma_c \over 2}$.
We see that the population of the level $c$ decays, to the first order of the
dipole interaction, on the same time scale $\gamma_{ab}^{-1}$ as $|a(t)|^2$.

This is an anti-Stokes process since the frequency $\omega_{cb}$ of the 
transition
to state $b$ is increased.  The rate of this anti-Stokes process is 
defined as the product of $|c(t)|^2$ and the rate $\gamma_{cb}$ for the 
transition
$c \rightarrow b$.

Calculating the total probability $N_{\gamma}^{aS}(T)$ of this process for the 
time
interval $T$ (duration of the laser pulse) smaller than the lifetime of the 
isomer 
$a$ ($T \ll \gamma_{ab}^{-1}$) and larger than the lifetime of the 
intermediate state $c$ (${1 \over \gamma_c} \ll T \ll {1 \over \gamma_{ab}}$), 
and dividing it by the probability $N_{\gamma}^{sp}(T)$ of the spontaneous 
decay
$a \rightarrow b$ during the same time interval, one obtains for the ratio
$$\eta = {N_{\gamma}^{aS}(T) \over N_{\gamma}^{sp}(T)}$$
the expression \cite{Bec84}

\begin{equation}
\eta = ({|V_{ac}|^2 \over 2\hbar})^2 {\gamma_{cb} \over \gamma_{ab}} 
{1 \over \Delta^2 + 
{\gamma_c^2 \over 4}}.
\end{equation}

For the matrix element $V_{ac}$ in (10), assuming that the nucleus is spinless,
 one finds

\begin{equation}
|V_{ac}|^2 = {1 \over 3}(eE_0)^2|\langle c|\vec{r}|a\rangle|^2
\end{equation}

\noindent
with $E_0^2 = {4\pi\hbar \over \lambda_L}\phi$, where $\phi$ is the photon 
flux and $\lambda_L$
is the wavelength of the radiation. 

The nuclear dipole matrix element $\langle c|\vec{r}|a\rangle$ can be expressed
 by the lifetime
$\gamma_{ca}^{-1}$ using a semiclassical calculation of Einstein's 
A-coefficient in the condition
of the thermal equilibrium \cite{Sar74}. The resulting relation coincides with
the result of the quantum 
mechanical Weisskopf-Wigner theory for the spontaneous decay \cite{Wei30}:

\begin{equation}
|V_{ac}|^2 = {\hbar \over 2\pi} {\lambda_{ca}^3 \over \lambda_L} \Gamma_{ca} 
\phi
\end{equation}

\noindent
with $\lambda_{ca} = {2\pi \hbar c \over E_c - E_a}$ and the width 
$\Gamma_{ca} = \hbar \gamma_{ca}$
for the transition $c \rightarrow a$.

\noindent
Finally, one obtains for $\eta$:

\begin{equation}
\eta = {\hbar \phi \over \Gamma_{ab}}({\lambda_{ca} \over \lambda_c})^3
{\lambda_L^2 \over 8\pi}
{\Gamma_{cb}\Gamma_{ca} \over (\hbar \Delta)^2 + {\Gamma_c^2 \over 4}}
\end{equation}

\noindent
where we introduced, instead of the rates $\gamma$, the corresponding widths, 
$\Gamma = \hbar\gamma$.

The factor $({\lambda_{ca} \over \lambda_c})^3$ in (13), which has been left
out of the first theoretical study
of this process, makes the experimental situation very difficult.  For 
instance, at laser photon energies of $\hbar\omega_L$ = 10 eV and a nuclear 
transition level difference $E_c - E_a$ = 100 keV
this factor is $10^{-12}$, which leads to very small $\eta$ values for 
moderate laser intensities.  The effect is noticeable only for small nuclear 
level spacings ($\lesssim$ 10-10$^{-2}$ eV).
The situation can be improved using resonant laser radiation with 
$\hbar\Delta \ll {\Gamma_c \over 2}$.  As can be seen from (10), in this case 
and under the conditions of moderate laser 
intensities ($|V_{ac}| < \Gamma_c$, but $|V_{ac}| \gtrsim 
(\Gamma_c\Gamma_{ab})^{1 \over 2}$, i.e., $I \gtrsim 10^{14}$ 
${W \over cm^2}$) $\eta$ can be of the order of unity. This means that the 
rates of two-step and one-step processes are equal.
For higher intensities ($|V_{ac}| > \Gamma_c$) the anti-Stokes process 
dominates.  This condition corresponds to $I \gtrsim 10^{19}$ ${W \over cm^2}$
(see \cite{Bec84} for details).  Under these conditions, Eqs. (7) must be 
solved precisely.  The resulting picture looks as follows
(again $\Delta \ll {\gamma_c \over 2}$, $\gamma_{ab} \ll \gamma_c$).
The system oscillates between the levels $a$ and $c$.  In contrast to the weak
field case ($|V_{ac}| \ll {\Gamma_c \over 2}$), where level $c$, as well as 
$a$,
decays with the slower rate $\gamma_{ab}$, now this laser-excited state decays
with the faster rate 
$\gamma \approx {\gamma_c \over 2}$.
The long lived isomer $a$ is depleted via $c$ to $b$ on a 
very short time scale. For the case of the isomeric state $^{111m}Pd$ one 
obtains $\gamma_c^{-1} \approx 1.5\cdot 10^{-11}$s ($\Gamma_{ca} = 2.3 \cdot 
10^{-6}$ eV (E1 transition), $\Gamma_{cb} = 4.3 \cdot 10^{-5}$
eV (E2 transition), $\Gamma_c \approx 4.5 \cdot 10^{-5}$ eV).
The condition $|V_{ac}| \gtrsim \Gamma_c$ corresponds to the intensity 
$I \gtrsim 1.3 \cdot 10^{19} {W \over cm^2}$.
The total probability for the anti-Stokes process in a very strong field 
approaches its maximal value

\begin{equation}
N_j^{aS}(T) = {\Gamma_{cb} \over 2\Gamma_c} {|V_{ac}|^2 \over \Gamma_{ab}
\Gamma_c + |V_{ac}|^2}
\approx {\Gamma_{cb} \over 2\Gamma_c},  \quad T\gg {1 \over \gamma_c}
\end{equation}

Hence, a pulse of $\approx$ 0.1 ns with the above intensity ensures that 
\underline{all} irradiated
isomeric nucleus decay.  The main experimental problem with the resonant 
excitation is the severe 
condition $\hbar|\Delta| \ll {\Gamma_c \over 2}$ which imposes very strong 
constraints on the
bandwidth ${\Delta\omega \over \omega}$ of the radiation necessary for $\eta$ 
to exhibit a resonant peak.  The associated requirements on the beam 
monochromaticity can be somewhat compensated for at the
expense of high intensity.  It is important to note that the theoretical 
treatment with only a single intermediate
atomic state is reliable only very close to the resonance.

Concluding this section, we notice that similar results to those obtained here
in the case of a strong radiation
field ($|V_{ac}| \gtrsim \Gamma_c$) apply to the nuclear $\beta$-decay 
\cite{WBE} where,
as mentioned above, $\gamma_{ca}$ is always much larger than $\gamma_{cb}$.  
For the $\gamma$-transitions
of nuclei one can find nuclei with $\gamma_{ca} \ll \gamma_{cb}$.  In this 
specific case the spontaneous decay of the
intermediate state $c$ tends not to go back to the isomeric state $a$ but 
leads into the final ground state $b$.

\section{Laser-assisted internal conversion}

\subsection{Theoretical background}
Internal conversion (IC) is a clear example where nuclear physics (deexcitation
of a nucleus) is merged with an atomic process (ejection of the electron from 
an atomic shell). Thus, in the 
spirit of the present review, it is quite appropriate to consider IC in 
detail, especially because the
theoretical approaches developed for studying laser assisted IC are common to
some other
nuclear-atomic processes where lasers play an important role. 

In this process, as already pointed out in the Introduction, the effective 
reduction of the multipolarity
of nuclear gamma-ray transitions plays a crucial role.  However, there exists 
another issue where lasers have a substantial influence: changing the atomic 
surroundings of a nucleus which affects IC (we 
already encountered this
kind of laser influence in Section 3 when we considered the role of lasers in 
the orbital electron capture).  
The influence of a laser field on IC has been investigated firstly just from
this point of view \cite{Bal82}.

For weakly bound electrons participating in the IC, the laser-assisted removal
of one of the electrons that significantly contribute to IC leads to a 
significant decrease of the coefficient of 
IC.\footnote{The coefficient of IC $\alpha_{IC}$ is defined as the ratio of
the probability
$W_{IC}$ for IC and the radiative transition probability $W_{\gamma}$ for a 
given nuclear transition
$E(M)L$ with electric (magnetic) radiation multipolarity $L$.  IC 
coefficients can be calculated by well developed methods \cite{IMB}.}
We consider here the more direct and efficient role of the intense laser 
radiation field on the IC 
process.  In general, if the IC process is non-zero in the laser free 
case ($\alpha_{IC}^{free}
\not= 0$) the change of $\alpha_{IC}$ due to the laser influence is small 
($\alpha_{IC}^{free} \gg 
\alpha_{IC}^{LA}$).  This is obvious, in principle, because the energy of the 
laser photons under study
($\lesssim 10$ eV) is generally small in comparison with the energy 
$E_{\gamma} = E_{IF} = E_I - E_F$ of 
the nuclear transition. (Below, we will consider the exception from this 
general rule in the case of 
low-lying nuclear levels.)  However, in nuclei where the energy difference 
$E_{IF} = E_{\gamma}$
is less than the magnitude of the electron binding energy $E_B$ ($\Delta = 
E_{\gamma} - |E_B| < 0$),
IC does not take place ($\alpha_{IC}^{free} = 0$).
If the photon energy of the laser beam $\hbar\omega_L$ is of the order of 
$|\Delta|$ the originally
forbidden IC process may occur via absorption of the necessary numbers of 
laser quanta.  Thus,
the electromagnetic radiation (laser) essentially accelerates the nuclear 
$\gamma$-decay rate and, more importantly, induces IC which 
is absent otherwise.  

We first consider some theoretical issues connected with the treatment of the 
laser-assisted IC.  The theoretical description of the 
electronic states under the joint action of the Coulomb potential of the 
nucleus and the intense 
radiation field is only approximate; no precise analytical solution is known.
We confine ourselves here to a simplified non-relativistic model 
\cite{Kal86,Kal88,Kal89,Zon90,Kar92} for the system ``nucleus +
electron + intense radiation field'' with Hamiltonian\footnote{There are 
several other approximate models in the literature, e.g., the  model where the
Coulomb potential of the nucleus is replaced by a
spherical harmonic oscillator \cite{Ber91}.} $H = H_o + H_I$, where $H_o = 
H_{oN} + H_{oe}$ is the sum
of the Hamiltonians for the unperturbed nucleus $H_{oN}$ (the explicit form of 
$H_{oN}$ does not enter in the
following treatment) and the \underline{single} electron $H_{oe}$ 
participating in IC, 

\begin{equation}
H_{oe} = -{Ze^2 \over R} + {1 \over 2m}(\vec{p} - {e \over c}\vec{A})^2
\end{equation}

\noindent
Here, $\vec{A}$ is the vector potential of the external laser field which 
will be specified later;
$\vec{R}$, $\vec{p}$, m denote electron coordinates, momentum and mass, 
respectively.
The interaction between the nucleus and electron is of the Coulomb type

\begin{equation}
H_I = -\sum_{p=1}^Z{e^2 \over |\vec{R} - \vec{x_p}|} + {Ze^2 \over R}
\end{equation}

\noindent
where $\vec{x_p}$ denotes the nuclear p-th proton radius vector. 
Relations (15) and (16) reflect a strong simplification, first of all, 
because the interaction between 
the electron and the intense radiation field is comparable to the binding 
energy of the electrons of the
atomic shell.  However, for the inner electrons (K, L-shells), the Coulomb 
potential of the nucleus
still dominates and the modification of the initial electron states by the 
laser field can be
treated perturbatively.  For the final (free) electrons one can use the 
well-known Volkov solution 
for the charged particle in the electromagnetic field \cite{Vol35}.  
Furthermore, since in this model 
one only uses (16) for the electron-nucleus interaction \cite{Bla79}, the 
photon exchange between 
the nucleus and the electron shell does not appear explicitly.  Such a 
simplified treatment 
emphasizes the corrections which at least are necessary to improve the theory 
of this process:\\
i) use of the relativistic Dirac Hamiltonian for the bound electronic state 
``dressed'' by the intense
radiation field.  This improvement requires mainly numerical calculations.\\
ii) taking into account the effect of the nuclear size, and\\
iii) the screening of the Coulomb potential of the nucleus by inner-shell 
electron clouds.\\
Item i) is most important and has no exact analytic solution (and the same 
holds for (15),(16)).

Thus, approximations are unavoidable, and we will be confined here to the 
simplifications
described above.  Since the final free electron is influenced by the intense 
laser field,
the corresponding wave function is based on Volkov's solution of the Dirac 
\cite{Vol35} and 
Klein-Gordon \cite{Bro64} equations for an electromagnetic plane wave with 
specially chosen initial
conditions.  

Volkov's paper has not received much attention for a long time,  presumably 
due to the absence of appropriate radiation sources at that time.  However, 
with the 
invention of the laser, the theoretical interest has been renewed, and 
different exact and approximate
solutions have been given with various choices of the initial conditions and 
different methods for the theoretical description
(classical relativistic and non-relativistic, quantum mechanical and quantum 
electrodynamical \cite{Nik64,Kel65,Nic66,Ehl78,Ber80}).  We give here the 
original Volkov solution
for the Klein-Gordon (KG) and Dirac(D) equations.  With the gauge for the 
radiation
field vector potential $A_0(x) = 0$, we have:

\begin{eqnarray}
\psi_{KG}(x) &=& e^{-ip\cdot x}exp[-{ie \over 2p \cdot q}\int_{-\infty}^{\beta}
d\beta' [2\vec{p}\vec{A}(\beta') - e\vec{A}^2(\beta')]] \notag \\
 \\
\psi_D(x) &=& \psi_{KG}(x)[1 + e{\vec{q}\cdot\vec{A} \over 2p\cdot q}]u(p), 
\notag
\end{eqnarray}

\noindent
where $p(\epsilon, \vec{p})$ is the four-momentum of the charged particle, 
$q(\omega, 
\vec{q})$ is one of the laser photons, $u(p)$ is a Dirac spinor, and
$\beta \equiv q\cdot x = \omega t - \vec{q}\cdot\vec{x}$.  The scalar product 
$p\cdot q = \epsilon\omega(1 - {\vec{v}\cdot\vec{q} \over |\vec{q}|})$ is in 
the nonrelativistic case ($v \ll 1$) simply given by $m\omega$, and 
$\psi_D(x) \approx \psi_{KG}(x)u(p)$, and finally $p \gg eA = {eE_0 \over 
\omega}$,
where $E_0$ is the amplitude of the laser electric field.

Under these conditions, the initial and final electronic states appear as two 
different approximations of the solution of the Hamiltonian (16):

\begin{equation}
\psi_i(\vec{R},t) = \phi_{nj\lambda}(\vec{R})e^{-iE_Bt \over \hbar}
exp({ie \over \hbar c}\vec{A}\cdot\vec{R})
\end{equation}

\begin{equation}
\psi_f = exp[-{i \over 2m\hbar}\int^t[\vec{p} - {e \over c}\vec{A}(t')]^2dt']
u_c^{(-)}(\vec{R})
\end{equation}

\noindent
Here $\phi_{nj\lambda}(\vec{R})$ is a hydrogen-type solution with quantum
numbers $n$, $j$, $\lambda$ and energy eigenvalue $E_B$ and 
$u_C^{(-)}(\vec{R})$ is an electron Coulomb wave function.  In relations (18) 
and (19) we restored, temporarily, $\hbar$ and $c$.

For the vector potential of the radiation field we write for the circular (c)
and linear ($\ell$) polarization:

\begin{equation}
A \begin{Sb} c \\ \ell \end{Sb} = \begin{pmatrix}a[\hat{e}_1\cos \beta + 
\hat{e}_2\sin \beta]\\
a\hat{e}_3\cos \beta\end{pmatrix}
\end{equation}

\noindent
The unit vectors $\hat{e}_1$, $\hat{e}_2$, $\hat{e}_3$, perpendicular to each
other, define the frame of reference, and $a = {cE_0 \over \omega}$.

Inspection of (18) and (19) shows that if we add to the condition
$p \gg {eE_0 \over \omega} = {eE_0\lambda_L \over 2\pi c}$
(where the momentum $p = \hbar k$ ($k$ - wave number)) the condition
${e \over \hbar c}\vec{A} \cdot \vec{R} \ll 1$, which is well satisfied for 
the inner shells, we achieve a great simplification: i) the space-dependent and
time-dependent parts of the S-matrix element of the process can be 
separated, ii) the space-dependent part is the same as in the laser-free case.

Notice that the first condition ($p \gg {eE_0\lambda_L \over 2\pi c}$) can be
rewritten as the condition that the kinetic energy of the final electron is 
larger than ${r_0I\lambda^2 \over 2\pi^2} \approx 10^{-14}I\lambda_L^2 \approx 
10^{-26}I\lambda_L^2$ eV where $r_0$ is the classical radius of the electron, 
and
the laser intensity $I$ is measured in ${W \over cm^2}$ and $\lambda_L$ in cm.
Thus, this number is extremely small ($\lambda_L \sim 10^{-5}$ cm) for any
reasonable intensity.  (See Appendix B for the dimensional analysis of 
scales characterizing the laser interaction with the electron.)
Using the above approximation, one is able to calculate the laser induced IC
coefficient $\alpha_{nj\lambda}^{L,las}$ for a transition of multipolarity L
and for an electronic state of quantum numbers $n$, $j$, $\lambda$.
The calculations include the point nucleus approximation ($|\vec{x}_p| < 
|\vec{R}|$),
the expansion ${1 \over |\vec{R} - \vec{x}_p|}$ of (16) in terms of spherical 
harmonics, and the Wigner-Eckart theorem which permits to express the matrix 
element of the
electric multipole moment of order $\ell,m$ through its reduced matrix element
between the nuclear states I and F \cite{Sob79}.
Because this reduced matrix element decreases rapidly with increasing $\ell$, 
we are 
confined to the lowest $\ell$ determined by the multipolarity L of the nuclear
transition, as usual.
The further evaluation involves the use of the addition theorem for spherical
harmonics and the orthogonality of the 3j symbols that enter here, and the 
averaging (and summing)
over the magnetic quantum numbers of the initial (final) nuclear and electronic
states, etc.

The approximation of neglecting the screening of the nuclear Coulomb potential
is improved somewhat by using instead of the charge $eZ$ the effective charge
$eZ_{eff}(n) = en({|E_B| \over R_y})^{1 \over 2}$, where $R_y = {e^2 \over 
2a_B}$, 
n is the principal quantum number of the electronic state with binding energy
$E_B$, and Bohr radius $a_B$.  The effective charge $eZ_{eff}$ corresponds to 
the charge of the hydrogen-like nucleus where the electron has the same binding
energy as in the real atom with the same quantum numbers.

As we emphasized above, we are interested in the situation where 
$\alpha_{IC}^{free}
= 0$, i.e., $\Delta < 0$.  For our purposes, of course, the most important 
case is the case near threshold: $\hbar\omega_{IF} \lesssim |E_B|$.  
The laser field must be so intense that the interaction energy of the electrons
with the laser field becomes comparable to the binding energy of the 
electrons in the shell.  This gives a condition for the laser field intensity

\begin{equation}
I \ge ({Z_{eff} \over n})^4[\hbar\omega_L(eV)]^26.31\cdot 10^8 \quad 
\text{in} {W \over cm^2}.
\end{equation}

\noindent
If one uses the small electron momentum assumption ${pna_B \over Z_{eff}} \ll 
1$, then $|\Delta| \ll |E_B|$.

The final result for $\alpha_{nj\lambda}^{L,las}$ depends on the following
parameters:
Nuclear (and atomic) parameters: $E_{\gamma} = E_I - E_F$, $L$ - multipolarity
of the nuclear $\gamma$-rays, $Z_{eff}$, and $\Delta = E_{\gamma} - |E_B|$.
The laser parameters enter in the following combinations: $r_0 = {\Delta \over
\hbar\omega_L}$, $I$, and $r = r_0 - {2e^2I \over m\hbar\omega_L^3}$.
The second term in the parameter $r$ multiplied by $\hbar\omega_L$ gives the
laser pondermotive potential which in some cases reduces the 
effectivity of the very high intensity $I$.

For the case $\Delta < 0$, the $\alpha_{nj\lambda}^{L,las}$ is expressed in 
terms of the threshold IC coefficient $\alpha_{nj\lambda}^{L,Th}$ which
is theoretically calculable.  Thus, if $\hbar\omega_L$ is comparable to
$|\Delta|$, then, for the laser-free forbidden case, IC can start after 
absorption of the corresponding number of laser photons demanded by 
energy conservation.  As a result, we have \cite{Kal88}

\begin{equation}
\alpha_{nj\lambda}^{L,las} = \alpha_{nj\lambda}^{L,Th}T_{\begin{pmatrix}c \\
l\end{pmatrix}},
\end{equation}

\noindent
where

\begin{equation}
T_{\begin{pmatrix}c \\ l\end{pmatrix}} = \sum_{N>-r}T_{\begin{pmatrix}c \\ 
l\end{pmatrix}}(b_N)
\end{equation}

\noindent
with index $c(l)$ corresponding to circular (linear) polarization.  $N$ in 
(26) is a positive integer.  Here $b_N = b_0(N + r)^{1 \over 2}$ with 

\begin{equation}
b_0 = {eE_0 \over m\omega_L^2}\sqrt{\hbar\omega_L \over R_y}{1 \over a_B} =
1.07 \cdot 10^{-6}({I \over (\hbar\omega_L)^3})^{1 \over 2}.
\end{equation}

As before we measure $I$ in ${W \over cm^2}$
and $\hbar\omega_L$ in eV.  The summation in (23) goes over the number of 
laser photons absorbed by the electronic shell.  In (23),

\begin{equation}
T_c(b_N) = {1 \over 2b_N}\int_0^{b_N}J_{2N}(x)dx,
\end{equation}

\noindent
where $J_{2N}(x)$ is the Bessel function.  The corresponding formula for
$T_{\ell}(b_N)$ includes the generalized Bessel functions $J_N(b,d)$:

\begin{equation}
T_{\ell}(b_N) = \int_0^1J_N^2(b_Nx, -{d \over 4})dx.
\end{equation}

For weak laser fields ($b_0 < 1$, or $I < 10^{12}(\hbar\omega)^3$),  applying
the small argument expression for the Bessel and generalized Bessel functions,
we obtain

\begin{equation}
T_c = ({b_{N_{min}} \over 2})^{2N_{min}}{(2N_{min})!! \over (N_{min}!)^2
(2N_{min} + 1)!!}
\end{equation}

\begin{equation}
T_{\ell} = ({b_{N_{min}} \over 2})^{2N_{min}}{1 \over (N_{min}!)^2(2N_{min} + 
1)},
\end{equation}

\noindent
where $N_{min}$ is the minimum number of laser photons necessary for the onset
of the originally forbidden IC process.  As $b_N^2 \sim I$, 
$T_{\left( \begin{smallmatrix}l \\ c\end{smallmatrix} \right)} \sim 
I^{N_{min}}$.

Curves exist in the literature \cite{Kal88} where the dependence of
$T(b_0, r_0)$ ((23), (25)) on the laser intensity is given for different 
materials 
and lasers with $I < 10^{14} {W \over cm^2}$.  Unfortunately, for higher 
$I$ ($\lesssim 10^{20} {W \over cm^2}$) such curves do not exist.  It is 
important to note that for larger values of $I$ ($> 10^{14-16} {W \over
cm^2}$) the hindering effect of the laser pondermotive potential
\cite{Ka892} must be taken into account.  This potential ${2e^2I \over 
m\hbar\omega_L^3}\cdot\hbar\omega_L$
($\approx 2 eV$ at $I = 10^{13} {W \over cm^2}$) that enters into the 
expression for
$r = r_0 - {2e^2I \over m\hbar\omega_L^3}$ leads to the necessity to 
increase $N_{min}$ in the summation of (23) and, thus, decreases the 
probability of the process. Fig 5 shows the above mentioned curves for 
$T(b_0, r_0)$.

The ``(anti) efficiency'' of the pondermotive forces depends on
the precise knowledge of the ionization thresholds of the electron states.  The
values for the ionization thresholds calculated in the nonrelativistic 
approximation can be changed appreciably in more precise calculations.  The 
actual
ionization threshold may be larger than the value suggested by the simplified
model.  Particularly, it was demonstrated \cite{Kul91} in a nonrelativistic
calculation for the hydrogen atom that a laser field can keep the electron in 
the atom even at very high intensity.  This phenomenon certainly needs more 
work to be
understood.  Recently, more refined calculations reduced the magnitude
in the laser intensity \cite{Dor93}.  Thus, a more detailed study of the 
ionization
process and its effects on the electronic states and their widths is required.

\subsection{Some examples: $N$ vs. $I$}
Consider as an example for the above considerations \cite{Kal89} an isomeric
E3 transition, $^{105m}Ag$, with $E_{IF} = 25.470$ keV.  The binding energy
$|E_B|$ for the K shell is 25.524 keV \cite{Led78}.  Thus, $\Delta$ = -44 eV,
and IC is forbidden for the laser-free case.  In the model considered above 
it is possible to calculate the laser threshold IC coefficient 
$\alpha_{1{1 \over 2}0}^{3Th} = 152$ \cite{Kal89}.

The comparison of this calculation with the general, relativistic IC theory 
\cite{Oco65} leads to an agreement within 20\%.

For a laser with $\hbar\omega_L = 6.42$ eV, i.e., for the number of 
laser photons $N=7$ necessary to start IC, and $I = 2.3\cdot 10^{14} 
{W \over cm^2} (b_0 = 1)$ one obtains
$\alpha_{1{1 \over 2}0}^{3Las} \approx 10^{-13}$, an unobservably small value.
For the hypothetical value of $\hbar\omega_L = 30$ eV which would lead to N 
between 1 and 2 and with $I = 2.4\cdot 10^{16}{W \over
cm^2}$ we would have $\alpha_{1{1 \over 2}0}^{3IC} = 0.16$ ($N=2$).

Consider now the two characteristic cases with $\Delta < 0$: the isomeric 
nuclei
$^{235m}U$ and $^{183m}W$, for which atomic and nuclear data exist
\cite{Led78}.  In Table II, the necessary data \cite{Led78} and the 
calculated results
for $\alpha_{nj\lambda}^L$ \cite{Kal88} are given.  For $^{235m}U$, $E_{IF} =
73.5$ eV; for $^{183m}W$, $E_{IF} = 544$ eV.  $\hbar\omega_L = 5$ eV was taken
for the laser photons.

\vspace{5mm}

\begin{center}
\begin{tabular}{|c|c|c|c|}
\hline
Atom & \multicolumn{2}{c|}{$^{235m}U$, $E_{IF}$ = 73.5 eV} & $^{183m}W$, 
$E_{IF}$ = 544 eV \\
\hline
Electron Shell & $5d_{3 \over 2}(O_4)$ & $5d_{5 \over 2}(O_5)$ & $4s_{1 \over 
2}(N_1)$ \\
\hline
Binding Energy $|E_B|$, eV & 105 & 96 & 592 \\
\hline
$\Delta = E_{IF}-|E_B|$, eV & -31.5 & -22.5 & -48 \\
\hline
$b=1.07\cdot 10^{-6}\cdot {n \over Z_{eff}}{I^{1 \over 2} \over 
(\hbar\omega_L)^{3 \over 2}}$ & 0.55 & 0.58 & 0.44 \\
\hline
$\beta_0$ & 5 & 5 & 9.5 \\
\hline
$I$, ${W \over cm^2}$ & $2.7 \cdot 10^{15}$ & $2.7 \cdot 10^{15}$ & 
$10^{16}$ \\ \hline
T & 0.27 & 1.2 & 3.6 \\
\hline
$\alpha_{nj\lambda}^{L,las}$ & $2.1 \cdot 10^{15}$ & $2.7 \cdot 10^{15}$ & 
$3.3 \cdot 10^2$ \\
\hline
\end{tabular}
\end{center}

\vspace{5mm}

\noindent Table II. Atomic and nuclear data and results of the calculations of
$\alpha_{nj\lambda}^{L,las}$ \cite{Kal88} for $\hbar\omega_L$ = 5 eV,
$I = \beta_0^2(\hbar\omega_L)^38.73\cdot 10^{11} {W \over cm^2}$, and
$\beta_0 = {eE_0 \over \omega_L}\sqrt{2 \over \hbar\omega_L\cdot mc^2}$.

\vspace{5mm}

The increase of the laser intensity (e.g., to $I = 10^{20-21}{W \over cm^2}$)
will lead, for the near threshold cases, $\Delta < 0$, to a drastic increase
of the laser assisted IC coefficients $\alpha_{nj\lambda}^{L,las}$.  The 
estimates lead to enormous values for $\alpha_{nj\lambda}^{Las}$.
Unfortunately, these estimates did not properly take into account the role
of laser pondermotive forces with their hindering role at very intense
fields. The detailed and accurate study at very high laser intensity 
will unavoidably demand the aid of modern, powerful computers.

Before turning to the experimental issues related to laser assisted
(and induced) IC we would like to mention that via the 
electron bridge mechanism it is also possible to accelerate the IC process
and to obtain higher values for the IC coefficients.  This subject will be 
considered in Section 5 which is devoted to the study of this mechanism.

\subsection{Experimental problems}
The main problems faced in experimental studies of laser assisted and induced 
IC are the production and collection of a sufficient amount of isomeric 
nuclei and their irradiation by intense ($I > 10^{14}{W \over cm^2}$) and 
relatively long laser pulses.  Typically, one experiences a high background of
ions and especially, electrons.  Due to the high electron
background, it may be more useful to detect the soft X-rays which are emitted
in the recombination processes when the vacancies left by the conversion 
electrons are occupied by electrons from higher shells, rather than to detect
the slow IC electrons.  

The number of emitted soft X-ray photons $N_{\gamma}$ can be estimated as

\begin{equation}
N_{\gamma} = \xi At{\alpha_{IC}^{las} \over \alpha_{IC}^T},
\end{equation}

\noindent
where $\alpha_{IC}^T$ is the \underline{total} laser free IC coefficient, 
$\alpha_{IC}^{las}$ is the laser induced IC coefficient for the shell under
study (see Table II), $A$ is the activity of the sample, $t$ is the total 
irradiation time (duration of the laser pulse in an ordinary case), and
$\xi$ is the efficiency of soft X-ray detection.

Taking, e.g., ${\alpha^{las} \over \alpha^T} \approx 0.2$ (for $^{183m}W$,
$\alpha^{las} = 3.3\cdot 10^2$, see Table II), $\xi = 10^{-2}$, $t 
\approx 10^{-9} s$, we obtain the necessary activity to obtain $N_{\gamma}
\approx 1$ in one laser pulse $A \approx 6\cdot 10^{11} Bq (15 Ci)$.
($1 Ci = 3.7\cdot 10^{10} Bq = 3.7\cdot 10^{10}$ disintegrations per second.)
The problem of producing a sample of low density but high activity can be
solved using a method similar to the one proposed in \cite{Let77} to sort
out the isomeric nuclei embedded in molecules, e.g., in our case, $WO_3$
\cite{Kal88}.  This method uses the fact that some isomeric nuclei have 
angular momenta higher than the ground state which leads to different molecular
spectra due to the difference in fine interactions.

The $^{183m}W$ isomer can be produced by thermal neutron capture:
$^{182}W(n,\gamma)$ $^{183m}W(\sigma \sim 2 \; mb)$ (the abundance of 
$^{182}W$
is 26.3\%).  Besides the fast isomer separation one needs a soft X-ray 
detection
method of high enough resolution in order to be able to select soft X-rays 
from the recombination of the atomic shells ($N_1$, $N_2$, $N_3$ in the case 
of $^{183m}W$).

Finally, the irradiation time $t$ may be increased to force the laser light
to run back and forth through the sample multiple times, while a laser active 
material compensates for the losses, thus keeping the intensity approximately 
constant.

Concluding this section, we notice that the specific IC process, where the 
shell electron remains in the discrete state, can be considered also from the
point of view of the Electron Bridge mechanism which we will study in the next
section.  Thus, the laser's role in the
elimination of the mismatch between atomic and nuclear level differences 
(resonant 
IC) will be clearly seen.

\section{Electron Bridge mechanism as a source of deexcitation of nuclei}

\subsection{Experimental observation}
A systematic way to use the electron shell as a mediator between the laser
beam and the nucleus for studying nuclear low-energy 
properties is via the so called Electron Bridge (EB) mechanism.

The idea was born long ago, in 1958, in the former USSR \cite{Kru58} and has
been developed more recently into a useful tool \cite{Kru68,Cra73}. The EB 
mechanism effectively transfers
the energy of a nuclear transition to the atomic shell which, passing
through excited intermediate state(s), emits monoenergetic $\gamma$-rays,
thus providing the nuclear deexcitation.  It is a 3rd order process with 
respect to the electromagnetic interaction (see Fig. 6 for the corresponding 
Feynman diagram).  The energy of the emitted $\gamma$-rays $E_{\gamma}$ is
$E_{IF} - (\epsilon_f - \epsilon_i)$ with $\epsilon_{i(f)} = -B_{i(f)}$, where
$B_{i(f)}$ is the binding energy of the electron in the atomic shell
$|i\rangle(|f\rangle)$.  When $B_i > B_f(\epsilon_f > \epsilon_i), E_{\gamma} <
E_{IF} \equiv E_I - E_F$, one obtains the so called ``Stokes line'' in the 
deexcitation spectrum.

The inverse EB (denoted as IEB below) mechanism corresponds to the analogous 
diagrams shown in 
Fig. 7 with the external radiation absorbed by the electron of the initially
unperturbed shell, thus, providing a way to excite nuclei.

The first experimental observation of the nuclear deexcitation via an EB
mechanism was done in 1985 by the Zagreb-Ottawa group \cite{Kek85}.
In \cite{Kek85} the deexcitation of the 30.7 keV (lifetime $T_{1 \over 2} \sim$
13.6 yr) isomeric level of $^{93}Nb$ was studied and 28.2 keV energy 
photons were observed as a result of the EB mechanism involving the initial
L-electron states (with binding energies $B_{L_1}$ = 2.675 keV, $B_{L_2}$ =
2.426 keV, $B_{L_3}$ = 2.368 keV) and the final state $N_5(B_{N_5} < 30 
\text{eV})$ or
higher electron states.  The higher L states were used instead of the K state,
since excitation of the K shell gives photons with energies below the niobium
K X-ray energy, and the background is too high.

It is necessary to point out that the EB processes are usually accompanied by 
processes connected with IC channels\footnote{As it was already stressed in the
pioneering papers \cite{Kru58} on the EB mechanism, such third order effects
in $\alpha$ should be substantial if the IC coefficients are very large.  That
reflects the close connection between the IC process (with an electron in the
continuum, or in the discrete state) and the EB mechanism where an additional 
photon is emitted.} which, as a second order process, is more probable, in 
general, than the 3rd order process under study (e.g., the external 
bremsstrahlung associated with IC of the 30.7 keV level).

There are also other contributions from the internal Compton scattering 
produced in the deexcitation of the initial nuclear level, etc. However, in 
\cite{Kek85}, these sources of background were found to be negligible.

Fig. 7 shows a spectrum of $\gamma$-rays after background subtraction.
The first theoretical estimate given by the authors of \cite{Kek85}
involves the numerical solution of the inhomogeneous Dirac equation for 
an electron in the nuclear Coulomb field and hydrogen-like atomic wave
functions. The ratio $\eta$ calculated for the probability $W_{\gamma}^{(3)}$ 
of the 3rd order transition based on the diagrams Fig. 6 and the probability
$W_{\gamma}^{(1)}$ of the direct nuclear deexcitation was found to be
$\eta_{th} = 
{W_{\gamma}^{(3)} \over W_{\gamma}^{(1)}} = 0.19$, whereas the experimental 
result was  
$\eta = 0.070 \pm 0.018$.  Later, more detailed theoretical calculations 
gave $\eta_{th}$ = 0.069. \cite{Kol90}

There exists a second experimental result \cite{Zhe88} based on the study of 
the 
decay of isomeric state $^{193m}Ir$ with energy 80.27 keV ($T_{1 \over 2}$ =
10 d). This isomeric state was obtained in the irradiation of a 99\% enriched
sample of $^{192}Os$ by thermal neutrons via the chain
$^{192}Os \stackrel{(n,\gamma)}{\rightarrow}$  $^{193}Os$ 
$\overset{\beta (0.3\%)}{\underset{T_{1 \over 2} = 30 h}{\rightarrow}}$ 
$^{193m}Ir$.

Experiments conducted under the conditions of high background associated with 
the
IC channel ($\approx 10^5W_{\gamma}^{(3)}$) gave for $\eta$ the value $\approx$
0.21.

Theoretical estimates for L electrons yielded $\eta$ = 0.18 which
should be taken only as the lower bound due to the neglect of the 
additional contributions from higher shells (M,N,...).
Estimates in \cite{Zhe88} give for $\eta$(M, N,...) $\sim$ 0.1-0.2.
Again, more precise calculations are in agreement with the experimental value
0.21 \cite{Kol90}.

Before turning to the role of lasers in the nuclear deexcitation process 
provided by the EB mechanism, it is worthwhile to note the following.  We 
have already
mentioned that the probability for IC increases with the effectiveness of the 
EB mechanism.
In turn, the existence of the EB channel leads to a significant decrease
in the IC coefficient with respect to its so-called ``tabulated'' value where
the EB mechanism is not taken into account.  For instance, for the isomeric 
state
$^{235m}U$, the ``tabulated'' IC coefficient ($\alpha_{IC}(E3) \approx 
10^{19}$)
is reduced by a factor of $\approx 10^5$ times \cite{Kol90}.

\subsection{The role of lasers in the EB mechanism of deexcitation of nuclei}
Now we will discuss the laser-assisted EB process where an electron of an atom 
containing an isomeric nucleus absorbs the nuclear transition photon
of energy $E_{\gamma} = E_{IF}$ and emits a number N of atomic transition
photons, thus eliminating the mismatch between $E_{IF}$ and
$\epsilon_{ni}$ (the energy difference between the intermediate and initial 
atomic
states). The resonantly excited electronic states finally decay by
X-ray emission.

The schematic diagram of Fig. 8 explains the process for the case of the 
emission of two
photons.  In this scheme, an electron from the initial shell
$|i\rangle$ ``dressed'' by the intense laser field absorbs a nuclear radiation
quantum with energy $E_{\gamma}$, emits two laser photons with energy
$2\hbar\omega_L$, and then occupies resonantly the intermediate state 
$|n\rangle$ with
energy $\epsilon_n$ ($\epsilon_n = -B_n$, where $B_n$ is the binding energy 
of the electron
in shell $|n\rangle$), and finally decays to the final state $|f\rangle$
(= $|i\rangle$ in this particular case) with emission of an X-ray photon with
energy $\hbar\omega_x$. Due to the dipole character of emission (absorption)
of the laser quanta ($\ell$ = 1), the laser also ``transfers'' angular 
momentum to 
effectively reduce the multipolarity L of the nuclear transition radiation to
L = 1 (L = 3 for the case of $^{235m}U$ considered below), thus enhancing
the probability of the process.

Fig. 9 presents the Feynman diagrams describing this process.  
At resonance, the
first diagram is dominating.  We have:

\begin{equation}
E_I - E_F = \epsilon_n - \epsilon_i + N\hbar\omega_L,
\end{equation}
or, for the detuning $\Delta$,

\begin{equation}
\Delta = E_I - E_F + \epsilon_i - \epsilon_n. 
\end{equation}

The well-known and theoretically studied case is the isomeric nucleus
$^{235m}U$ with energy $E_I$ = 73.5 eV (angular momentum $J_I = {1 \over 2}^+$,
  $T_{1 \over 2} = 26$ min)
which has an E3 transition to the ground state F with $J_F = {7 \over 2}^-$.

Among the numerous electronic orbits of the uranium atom we choose the orbit 
$6s_{1 \over 2}$ with binding
energy 71 eV ($\epsilon_i$ = -71 eV), which is closest to $E_{\gamma}$, as 
the initial 
electronic shell $|i\rangle$.
We describe the complex electronic structure of the uranium atom in 
the one-electron approximation, i.e., we neglect
the splitting of shells with definite principal quantum number.  At resonance 
this approximation is appropriate.  

As an intermediate atomic shell, we select the $8s_{1 \over 2}$ shell with 
binding energy 2.14 eV
($\epsilon_n$ = -2.14 eV).  Thus, the detuning $\Delta$ is 4.64 eV.  The 
resonance will be achieved 
by two (four) laser photon emission with energy 2.32 (1.16) eV, respectively. 
The final atomic
one-electron state $|f\rangle$ is supposed to be the same as $|i\rangle$ 
\cite{Kal91}.  All electronic
states ($|i\rangle$, $|n\rangle$, $|f\rangle$) are ``dressed'' states.

Due to the laser, not only the resonance condition is fulfilled, but the 
originally E3 $\gamma$ transition is 
converted into the emission of an E1 X-ray photon, thus increasing the rate of
the X-ray emission significantly.
The laser beam makes it possible to avoid the selection rules of X-ray 
emissions which are acting in the laser free case.

The reader may notice the arbitrariness in the selection of the intermediate 
state $|n\rangle$ and its
energy $E_n$.  However, the (assumed) value of $E_n$ is not so important for 
the phenomenon under
study from the point of view that it is a small change, whereas the change of 
the \underline{order} of the
resonance (i.e., the number N of laser photons needed to fulfill the resonance
conditions, $\Delta = N\hbar\omega_L$)
is crucial.

\subsection{Theoretical considerations}
The theoretical study of EB processes is analogous to the calculations of 
the laser assisted IC which
were presented in Section 4.  The theoretical framework was developed in the 
papers \cite{Hin81,Kal91,Ka912,Zon90,Kar92,Str91}.

The study \cite{Kal91,Ka912} is based on the Hamiltonians $H_0$ and 
$H_I$ (15), (16), and we take as 
substitute for (15) the dipole form (1): The interaction of the linearly 
polarized laser field with the
electron is described classically by a Hamiltonian of type (1) $H_{Le} = 
-e\vec{r}\cdot\vec{E}_L(t)$
and the interaction of the emitted (quantized) X-ray field is described by 
the same kind of Hamiltonian
$H_{ex} = -e\vec{r}\cdot\vec{E}_x(t)$, where

\begin{equation}
\vec{E}_L(t) = \vec{E}_0\hat{\epsilon}_3\cos\omega_Lt
\end{equation}

\begin{equation}
\vec{E}_x(t) = i\sum_{\omega_x,\epsilon}({2\pi\omega_x \over V})^{1 \over 2}
\vec{\epsilon}
(ae^{-i\omega_xt} - a^+e^{i\omega_xt})
\end{equation}

\noindent
Here $\omega_x$ and $\vec{\epsilon}$ are the frequency and unit polarization 
vector of the emitted
radiation, $a$ and $a^+$ are the annihilation and creation operators, and $V$ 
is the normalization volume, respectively.

Again, as before, the influence of the other electrons is taken into account 
by the use of the effective
nuclear charge $Z_{eff}(n)$ in the one-electron wave functions.  As before, 
due to the expression (16) that describes the electron-nucleus interaction, 
photon exchange between the nucleus and the electron shell does not appear 
\cite{Bla79}.

In the treatment of the joint influence of the nuclear Coulomb and laser 
fields on the dressed electron in the subshell $|n\rangle$, it is convenient 
to use parabolic coordinates \cite{Kal91} for the shell
of principle quantum number $n$, $n = n_1 + n_2 + |m| + 1$, here $n_1$, $n_2$ 
are the parabolic quantum
numbers \cite{Lan77}, and $m$ is the magnetic quantum number.

The wave function describing the ``dressed'' electron state can approximately 
be written as \cite{Kal93}

\begin{equation}
\psi(n, n_1, n_2, m) = \phi_{n_1n_2m}\sum_{N=-\infty}^{+\infty}J_N
(\lambda_{n_1n_2})
e^{-i(E_n + N\hbar\omega_L){t \over \hbar}},
\end{equation}

\noindent
where $\phi_{n_1n_2m}$ is a hydrogen-type solution in parabolic coordinates, 
$J_N$ is a Bessel function
of the first kind, $N$ is the number of absorbed or emitted laser photons, 
$E_n = \epsilon_n - 
{i\Gamma_n \over 2}$ is the complex energy of the intermediate electronic 
state $|n\rangle$ with its
energy $\epsilon_n$ and width $\Gamma_n$, and finally the argument of the 
Bessel function is 

\begin{equation}
\lambda_{n_1n_2} = {3n(n_1 - n_2)E_0ea_B \over Z_{eff}\hbar\omega_L}.
\end{equation}

\noindent
This solution is used as a basis for the construction of the electron Greens 
function and S-matrix for the process.

The further procedure is analogous to the one we described for IC (section 4).
The initial and final state
is taken as $is$- and $fs$-type, respectively, where $i(f)$ is the principal 
quantum number for the $|i\rangle(|f\rangle)$ state
$R_{i(f)0}(r)Y_{00}$. Here, $R_{i(f)0}$ is the nonrelativistic radial part of 
the hydrogen-type solution
of principal quantum number $i(f)$ and orbital angular momentum $\ell$ = 0, 
and $Y_{00}$ is the corresponding
spherical harmonics.  $Z_{eff}$ for the $|i\rangle$ state (n = 6) is 
equal to $Z_{eff}$(6) = 13.71,
and for the $|n\rangle$ state (n = 8), $Z_{eff}$ (8) = 3.17.

As a final result of this procedure, we obtain for the ratio 
$\eta = {W_{fi}^{las} \over W_{fi}^{sp}}$ of the probability $W_{fi}^{las}$ of
the laser assisted resonant EB mechanism and the probability $W_{fi}^{sp}$ of 
the spontaneous $\gamma$ decay \cite{Bla79} the following expression:

\begin{equation}
\eta = {2e^2 \over 9}{[(2L - 1)!!]^2 \over (L + 1)(2L + 1)}{|J_{ni}|^2
|I_{L,ni}|^2 \over
(\Delta - N\hbar\omega_L)^2 + {\Gamma_n^2 \over 4}}[{c \over 
\omega_{IF}}]^{2L - 2}F,
\end{equation}

\noindent
where $\Gamma_n = \Gamma_{n0} + \Gamma_{nf}$, and $\Gamma_{n0}$ is the natural 
linewidth of the state
$|n\rangle$ for the laser-free case and $\Gamma_{nf}$ is the laser field 
contribution to the power broadened linewidth,

\begin{eqnarray}
J_{ni} = \int dr r^3R_{i0}(Z_{eff}(i),r)R_{n1}(Z_{eff}(n),r) \notag \\
 \\
I_{L,ni} = \int dr r^{1-L}R_{i0}(Z_{eff}(i),r)R_{nL}(Z_{eff}(n),r). \notag
\end{eqnarray}

\noindent
For given values of $i$=6, $n$ = 8, $L$ = 3, $Z_{eff}(i)$, $Z_{eff}(n)$ we 
have for 
$J_{ni}$ and $I_{L,ni}$
$$J_{86} = -0.51a_B, I_{3,86} = 4.4\cdot 10^{-5}{1 \over a_B^4}$$.

The quantity F in (36) is 

\begin{equation}
F = \sum_K({\omega_{xk} \over \omega_{IF}})^3F_K = \sum_Kf_K,  
\end{equation}

\noindent
where $\omega_{xK} = \omega_{IF} - N\omega_L + K\omega_L$, and $K$ is an 
integer.
The quantities $f_K$ are quadratic combinations of products of the Bessel 
function $J_K(\lambda_{n_1n_2})
\cdot J_K(\lambda_{n_1n_2})$ with the proper Clebsh-Gordon coefficients.  
$f_K$ are maximal at $K = 0,\pm 1$.

Figs, 10,11 show the intensity I, and the K dependence of $F_K(f_K)$ 
\cite{Ka912}.

At the resonance ($\Delta = N\hbar\omega$, $N$ = 2 or 4, $\hbar\omega_L$ = 
2.32 eV or 1.16 eV, respectively)
and a intensity $I \approx 10^{12}{W \over cm^2}$, with an estimated value of 
$\Gamma_n \approx 10^{-5}$
eV (from a $2p \rightarrow 1s$ transition of the same energy) we obtain for 
the $^{235}U$ isomeric state \cite{Kal91}

\begin{equation}
\eta \approx 6.7\cdot 10^9.
\end{equation}

If one takes into account the fact that the intense laser broadens the 
linewidth $\Gamma_{nf}$ \cite{Ka912},
the magnitude of $\eta$ decreases but still remains of the order of $10^9$, 
leading to expectations
that the affect is accessible to experimental study, though the IC coefficient
 is much, much higher.  We 
note that above we concentrated on the special case of the bound-bound 
atomic transitions where the final
electron shell is the same as the initial one ($|i\rangle = |f\rangle$).  
In this case the IC strongly dominates
over the EB process.   

Concluding this part of our consideration of the EB processes and the laser's 
role
in the acceleration of the radiative deexcitation of the nucleus, we 
emphasize that the above example
clearly shows again the advantage of the use of intense lasers in this field:\\
-due to the high intensity, lasers provide the fulfillment of the resonant 
conditions ensured by the
multiphoton absorption (emission).\\
-multipolar nuclear $\gamma$-transitions are converted into the emission 
(absorption) of electric
dipole radiation.  Laser radiation transfers not only the energy to the shell 
to eliminate the mismatch
between the atomic and nuclear transition energies, but also ``transfers'' 
angular momentum to provide
the dipole character of the final radiation, which leads to an enhancement of 
the probability of X-ray radiation by the final electron.

Turning now to the case of the resonant atomic transitions 
$|i\rangle \rightarrow |n\rangle \rightarrow
|f\rangle \not= |i\rangle$, we use the simple estimates based on the 
argument that the EB mechanism can
be considered as a kind of IC process where the shell electron, 
absorbing the $\gamma$-radiation
of the nucleus, undergoes a transition not to the continuous spectrum but 
rather to the discrete spectrum.  We shall refer to this conversion mechanism 
as ``discrete 
conversion''. Just as lasers assist the ordinary IC process, we also can  
speak here 
about the ``laser assisted discrete internal conversion''. In the case of 
tuning by laser quanta to remove
the mismatch between nuclear and atomic transition frequencies, we refer to 
the ``resonant discrete internal conversion.''

This way of treating the acceleration of the nuclear deexcitation process by 
resonant laser photons
presents an alternative method of the theoretical study of the EB mechanism 
\cite{Zon90,Kar92}.  It is quite
useful since it provides parallels to the well developed and accepted methods 
used in studies of IC phenomena.

The extension of the theory of IC (and ``resonant'' IC) to the discrete case 
introduces a new issue since
the formal determination of the ``discrete IC'' coefficient $\alpha_d$ leads 
to a quantity with the dimension
of energy, so that the physical interpretation of the IC coefficient as the 
ratio of conversion and radiative
transition probabilities is lost.  The reason for that is clear since the 
normalizations of the final 
state wave functions for the continuous and discrete states are different.

This view of the EB process as an IC process for discrete electron final 
state gives for the discrete
conversion coefficient $\eta^d = {W_{IC}^d \over W_{\gamma}^{(1)}}$ defined 
as the ratio of the probabilities
for discrete conversion $W_{IC}^d$ and direct nuclear deexcitation 
$W_{\gamma}^{(1)}$ introduced
previously, the simple expression which does not depend on the nuclear 
matrix element:

\begin{equation}
\eta^d = {\alpha_d(M(E)L)\Gamma_n \over 2\pi(\Delta^2 + {\Gamma_n^2 \over 4})},
\end{equation}

\noindent
where $\alpha_d$ for the $E$ type transition (in our case of the $^{235}U$
nucleus) has the form:

\begin{equation}
\alpha_d(EL) \approx {\alpha L \over L + 1}(2j + 1)[(2L - 1)!!{\mathbf M}]^2
\omega_{IF}^{-2L-1},
\end{equation}

\noindent
and $\mathbf{M}$ is the matrix element for the atomic transition $|i\rangle 
\rightarrow |f\rangle$

\begin{equation}
{\mathbf M} = \langle f|r^{-L-1}|i\rangle = \int_0^{\infty}g_i(r)g_f(r)
r^{1 - L} dr,
\end{equation}

\noindent
where $g_{i(f)}$ is the nonrelativistic (large component) radial wave function
for the electron in the
$|i\rangle(|f\rangle)$ state.  For the $^{235m}U$ isomeric state, in the laser
free case of a large 
resonant defect, $\Delta \gg \Gamma_n$, we obtain $\eta^d \approx 
{\alpha_d(E3)\Gamma_n \over 2\pi\Delta^2} \approx
2\cdot 10^{12}$.
Despite the fact that the EB mechanism (laser free case) is more probable than
the direct radiative transition by
twelve orders of magnitude, it is much smaller than the IC coefficient 
$\alpha_{IC}(E3)$($\approx 10^{19}$
for $^{235}U$).

In the case of discrete conversion in a laser field with resonant 
tuning, 
$\Delta - \hbar\omega_L \ll \Gamma_n$, one obtains for $\eta_L^d$ \cite{Zon90}:

\begin{equation}
\eta_L^d = {2\alpha_d(E3)|\beta|^2 \over \pi\Gamma_n},
\end{equation}

\noindent
where $\beta$ determines the matrix element for the transition from the 
intermediate atomic
shell $|n\rangle$ (here $5f_{5 \over 2}$) to the final state $|f\rangle$ 
(taken here as $6d_{3 \over 2}$)
induced by the interaction of the laser field with the electron 
$H_{Le} = -e\vec{E}\vec{r}$:

\begin{equation}
\beta = {1 \over \Delta}\langle n|H_{Le}|f\rangle
\end{equation}

\noindent
(In contrast to the previous case, here $|i\rangle \not= |f\rangle$:
$6s_{1 \over 2} \rightarrow 5f_{5 \over 2} \rightarrow 6d_{3 \over 2}$, see 
Fig. 12). The
parameter $\beta$ can be calculated in the quasiclassical approximation and 
was found to be $\sim 0.3 a_B$
for the above $|n\rangle$ and $|f\rangle$ states.  This gives 
$\eta_L^d \approx 7 \cdot 10^{18}$ for the moderate laser intensities $I \sim 
10^{12} {W \over cm^2}$, 
which is comparable with the IC coefficient for the outer atomic shells.  This
indicates a considerable reduction in the
lifetime of the isomeric state $^{235}U$ due to the influence of laser 
radiation.  In reality, this reduction must be multiplied by 
$\Delta t \cdot \nu$, where  $\Delta t$ is the duration of the laser pulse 
and $\nu$ 
is the pulse repetition rate.  In practice, this factor is much less than 
unity.  

For higher laser intensities ($I \gtrsim 10^{14} {W \over cm^2}$) the 
estimates are complicated by the
formation of multiply charged uranium ions due to multiphoton ionization. This
leads to
additional (ionization) broadening of the $|n\rangle$ and $|f\rangle$ atomic 
levels which is comparable
with the radiative width used above for $\Gamma_n$ ($4\cdot 10^{10}s^{-1}$).
At these intensities, $\eta^d$ is found to be $\sim 7 \cdot 10^{20}$ 
\cite{Zon90}.

As claimed in \cite{Kek85}, due to the fact that the ionization of the atom 
increases the localization of the 
electron wave function near the nucleus of the ion as compared to the case of 
a neutral atom (a well-known effect
in the theory of IC), the matrix element ${\mathbf M}$ in (42) increased 
significantly. This leads to an
the additional increase of $\alpha_d$ in (43), and, therefore, of $\eta_L^d$.
Since the 
resulting value for $\eta_L^d$ is expected to be much greater than 
$\alpha_{IC}$ for the outer shells, the 
lifetime of the isomeric state $^{235m}U$ would be reduced by two or three 
orders of magnitude for laser
intensities $I \gtrsim 10^{14} {W \over cm^2}$.

At present, it is not clear how to extrapolate these results to higher 
intensities because of the difficulties associated with
the theoretical estimate of the widths $\Gamma_n$ of the intermediate atomic 
states due to the
copious formation of multiply charged ions by very strong laser fields.  One 
expects that at high
intensities, when the ionization broadening of the electronic states is higher
than their radiative widths and where the broadening
increases faster than linearly with $I$, $\eta^d$ will be reduced. 

The estimates of this Section demonstrate that, in principle, the decay rate of
the isomer $^{235m}U$ to the 
ground state can be controlled by the use of intense laser radiation.

Particularly, the EB process or the ``resonance'' discrete and continuous 
conversion can be used as a method of  
producing vacancies in the atomic shells.  The coherent radiation resulting 
from filling of these vacancies, 
interesting enough itself, can be used for the detection and study of the 
resonant EB processes.

\subsection{Inverse Electron Bridge mechanism of nuclear excitation}
The discussions presented in the preceding subsection in connection with the 
``resonant discrete'' IC
providing a control mechanism for the decay rate of nuclear isomers   
apply also to the reverse process: a nucleus is excited by an 
electron transition when the laser 
radiation resonantly eliminates the mismatch between the electron level energy
difference $\epsilon_{fi} =
|\epsilon_f - \epsilon_i|$ and the nuclear level energy difference
$E_{FI} = E_F - E_I$.

The situation where the external radiation (e.g., X-rays) transfers, through 
the excitation of the 
atomic shell, energy to a nucleus which is initially in its ground state is
called inverse electron bridge (IEB) 
mechanism. If the energy of the 
electron transition $\epsilon_{fi}$ is close to the energy of the nuclear 
transition $E_{FI}$, then the 
resonantly enhanced excitation of a nucleus can be achieved by the absorption 
or emission -depending on whether
$E_{FI}$ or $\epsilon_{fi}$ is larger - of a number of laser photons in 
addition to an X-ray photon.  Again,
the excitation of the electronic shell by the combined action of external 
radiation and resonant laser fields
(first stage) can be more effective than the direct excitation of the nucleus 
by external radiation
due to the dipole character of the interaction of the electron shell with the 
long wavelength radiation field. 
In the second stage the input energy is converted into  
energy of high multipolarity 
radiation needed for nuclear excitation.  Of course, the applied laser 
field must be tuned to resonance to fulfill
the requirement of energy conservation in this stage.  The level scheme (a) 
and corresponding Feynman 
diagram (b) are presented in Fig. 13.

The (approximate) theoretical treatment of this process is very similar to the
laser-assisted nuclear deexcitation considered
above.  This is reflected in the use of the Hamiltonian (16), expressions of 
the type (1) for the 
Hamiltonians $H_{eL}$, and $H_{ex}$ describing the interactions of the 
electron with the laser field and the X-ray field,
respectively, Eqs. (32), (33), and Eq.(34) for the wave function of the 
electron state ``dressed'' by the
intense laser field, as well as (35), etc.
All these approximations mean that the diagram in Fig. 13(b) must be replaced 
by the diagram in Fig. 14, where only 
the intermediate atomic state is dressed according to the sum of the diagrams 
shown in Fig. 15 and is described by the wave
function (34), although, strictly speaking, both $|i\rangle$ and $|f\rangle$ 
states have to be considered as dressed states, too. 

The dressed intermediate electronic state is able to absorb the laser photons
tuned to resonance before the nuclear excitation takes place.

\subsection{Some estimates}
We omit detailed calculations which are similar to the previous one for 
obtaining formulae (36), (37), and (38).

Let us take again, as an example, the $^{235m}U$ isomeric 
state \cite{JAB,Led78,Shm93} with $E_{IF}$ = 73.5 eV, $L$ = 3 and
$O_4(5d_{3 \over 2})$($E_B = 105$ eV) and $P_3(6p_{3 \over 2})$($E_B = 
32.3$ eV) shells as $|i\rangle$ and $|f\rangle$
states.  
The intermediate state $|n\rangle$ of binding energy $E_B$ = 2.14 eV can 
be excited from the initial 
$O_4$ shell by the absorption of a soft X-ray photon with energy 
$\hbar\omega_x$ = 103 eV if no additional
laser photon is absorbed (N = 0 case).  The effective charges and the principal
 and angular momentum quantum numbers
of the electronic states are:\\
$|i\rangle : Z_{eff} = 13.89, n_i = 5, \ell_i = 2(j_i = {3 \over 2})$, ($E_B =
105$ eV)\\
$|n\rangle : Z_{eff} = 3.173, n = 8, \ell = 3 \text{~or~} 1$, ($E_B = 2.14$ 
eV)\\
$|f\rangle : Z_{eff} = 9.245, n_f = 6, \ell_f = 1(j_f = {3 \over 2})$, ($E_B =
32.3$ eV).\\
The energy mismatch is 0.8 eV, i.e., tuning of the laser photon around 
$\hbar\omega_L \approx 1$ eV (N = 1) will result in a
resonant excitation of $^{235}U$ from the ground state to its isomeric state 
$^{235m}U$.

Using an intensity of the laser field of $I = 10^{11} {W \over cm^2}$ (to 
avoid the power broadening of the atomic
widths) one obtains for $\eta_{\gamma}$ defined as the ratio of the yield for 
nuclear excitation by laser-assisted
IEB process to the yield of direct nuclear excitation by $\gamma$-ray 
absorption, the value
$\eta_{\gamma} \approx 3\cdot 10^{12}$ \cite{Kal93}.

We conclude with the statement that the combined application of an X-ray 
source and an intense laser beam
for nuclear excitation can be more effective than the direct 
$\gamma$-excitation if among the atomic 
electron shells there are two with an energy difference close to the 
nuclear excitation energy. 
The exact tuning of the electron transition to resonance is established by 
applying an intense laser of appropriate
photon energy and intensity.  Under these conditions, the laser-assisted IEB 
mechanism opens a realistic method for nuclear
excitation with energy differences $E_{IF}$ between the excited and ground 
states of the order of several keV.

Concluding this Section, we would like, for the sake of completeness, to 
mention that there exists an 
alternative way to induce the nuclear transitions by producing electron 
holes in the atomic shells by a laser 
induced plasma \cite{Mor73}.  For the example of $^{235}U$ considered above, 
for instance, a hole is produced in the
$5d_{3 \over 2}$ shell, then the electron of the above mentioned 
$6p_{3 \over 2}$ state transits to this hole
($6p_{3 \over 2} (\epsilon_i = -32.5 eV) \rightarrow 5d_{3 \over 2} 
(\epsilon_f = -103.1 eV)$); the emitted photon excites the ground state of 
$^{235}U({7 \over 2})^-$ to the isomeric state with $J^P = {1 \over 2}^+$
(73.5 eV, $T_{1 \over 2}$ = 26 min.).
The laser produced plasma has ahigh electron density ($\gtrsim 10^{19} 
cm^{-3}$).

The calculations are analogous to the ones developed in this section for the 
study of the laser-assisted
IEB mechanism as a means for exciting nuclei. However, the difference is that 
the photon spectrum of the self
radiation of the plasma is continuous, with a Planck frequency distribution:

\begin{equation}
n(\omega_{Pl}) = {2 \over \pi}{\omega_{Pl}^2 \over exp({\omega_{Pl}
 \over kT}) - 1},
\end{equation}

\noindent
where $T$ is the plasma temperature and $k$ is the Boltzmann constant.

The excitation of nuclei in a plasma via the IEB mechanism was investigated 
theoretically in \cite{Tka90}.

Two cases are interesting here.  If the plasma temperature $T$ is so high 
that complete ionization of the atomic levels that normally would participate
in the IC process, takes place, then the IEB will occur through the discrete 
part of the 
spectrum.  If, however, $T$ is too low to provide the complete ionization of 
such levels (i.e., $E_{IF}\gg kT$),
then the IEB goes mainly through the continuous part of the spectrum.  
In \cite{Tka90}, the ratios of the 
efficiencies for the excitation of nuclei by the plasma photons through the 
IEB mechanism and by thermal radiation \cite{Aru89} were calculated for both 
temperature cases.

For the example of the uranium nucleus ($6p_{3 \over 2} 
\rightarrow 5f_{3 \over 2}$ transition) and in the case of the
transitions through discrete atomic levels ($E_{IF} \ll kT$), this 
ratio is $\approx 10^{17}$ \cite{Tka90}.

This effect was observed in \cite{Iza79} for $CO_2$ ($\lambda_L = 10.6 \mu m$)
and Nd ($\lambda = 1.06 \mu m$) lasers
where the production of the isomers of the $^{235}U$ nucleus was observed by 
the detection of IC electrons from 
isomeric levels.  The increase ($\sim 10^2$ times) of the isomer yield 
obtained with the shorter wavelength laser was observed, too.

In \cite{Gol76,Go762,Gol81} the capture mechanisof electrons from the 
plasma continuum by 
an outer electronic shell is predicted.

\section{Lasers in the study of anomalously low-lying isomeric nuclear 
states: ($^{229m}Th$)}

\subsection{Introductory remarks}
In this section we use the ideas and theoretical treatments which were 
already discussed and applied in some detail in the previous sections.  We 
think that the specific role of the phenomenon of anomalously low-lying 
nuclear state ($\approx$ several eV) justifies a detailed discussion.  The 
problem has a relatively long history.

In 1989 the results of the first relatively accurate experimental 
determination of energies and intensities of $\gamma$
transitions populating the ground state and first excited state of $^{229}Th$ 
obtained from the $\alpha$
decay of $^{233}U$ ($^{233}U \rightarrow ^{229}Th + \alpha$) were 
reported \cite{Rei90}.
Earlier, in the study of the rotational-band structure of $^{229}Th$ it was 
concluded \cite{Kro76} that the excited
${3 \over 2}^+[631]$\footnote{Set $[Nn_Z\Lambda]$ are the Nilsson model quantum
numbers} state is located quite close
(within 100 eV) to the $^{229}Th$ ground state (${5 \over 2}^+[633]$).  This 
statement was indirectly
supported by the study of the $^{230}Th(d,t)$ $^{229}Th$ reaction 
\cite{Tot78,Bur90}.  Later it was established that 
the energy difference between the isomeric state and ground state of 
$^{229}Th$ does not exceed 10 eV \cite{Ako89}.
In the paper quoted above \cite{Rei90} this upper limit was reduced to 5 eV 
((-1 $\pm$ 4) eV).
Unfortunately, the techniques used in these investigations did not make it 
possible to determine the energy of the isomeric
state $E_{is}$ more accurately or to measure its half-life time.
Furthermore, it is worth  mentioning that from these studies it can not 
definitely be
excluded that the state ${3 \over 2}^+[631]$ is the ground state
of $^{229}Th$ as opposed to the state ${5 \over 2}^+[633]$ \cite{Kar92,Rei90}.

In a recent paper \cite{Hel94}, in an attempt to improve the value for the 
energy of the $^{229}Th$ isomeric level,
the authors of \cite{Rei90} remeasured the energies of a number of $\gamma$ 
rays associated with the $\gamma$-decay of $^{233}U$.   
Compared with their earlier study, they considered
more $\gamma$ rays in the $^{229}Th$ spectrum,  used more well-measured energy
calibration and reference lines, and more
detectors with better energy resolution. They were able to more closely match 
the counting rates in the $\gamma$ ray peaks
whose relative energies had to be measured, and to reduce systematic errors.  
More than 111 $\gamma$-ray spectra
were measured, and a value of 3.5 $\pm$ 1.0 eV was determined for the energy 
of the low-lying isomeric level
of $^{229}Th$.  In these undoubtedly improved measurements, again, it was 
assumed that the ${3 \over 2}^+$ state lies above the ${5 \over 2}^+$ state.

The importance of the existence of low-lying isomeric states, unusually low 
on a nuclear energy scale, is obvious.  This includes not only nuclear physics
itself, but also optics, solid-state physics, lasers, plasma, etc.
For example, considering the great sensitivity of these levels to the 
electronic stucture, the lifetimes of such states are
expected to depend on the chemical and physical environment in which these 
isomers are embedded.  Thus, the knowledge of the life-time, e.g., of 
$^{229m}Th$ in different
chemical and physical environments could provide valuable information for 
atomic and condensed matter physics.
It is important to emphasize that closely spaced levels with energy spacing of
several tens eV are 
encountered quite frequently at excitation energies of the order of 
$\sim 10^2$ eV.  However, they have 
vanishingly low probabilities in comparison with the background connected with
decays to low-lying states or to the   
ground state.  These transitions can be studied only in the isomer decays of 
\underline{low-lying}  levels. The 
frequently considered example of $^{235}U$ is typical from this point of view.
However, its isomer, as we discussed already,
decays via the IC channel, despite the fact that the efficiency of the EB (or 
IEB) processes is very high for $^{235}U$ ($\eta \gg 1$;
see Section 5) it is small on the scale of the IC process, which 
dominates here($\alpha_{IC} \approx
10^{19}$).  For other examples considered above, (e.g., $^{93m}Nb$ or 
$^{193m}Ir$), $\eta < 1$ and again IC is the dominant process.  From this 
point of view, the situation of $^{229}Th$ is
entirely different: the ionization potential of the thorium atom is 6.08 eV, 
so that a transition of the isomeric
nucleus $^{229m}Th$ to the ground state via IC with a final unbound electron 
is impossible (at least, for an isolated atom). 

These considerations are the basis for the high interest in laser radiation as
a means to accurately study the properties of $^{229}Th$
(energy level differences, lifetime of isomeric nucleus, etc.).

\subsection{Deexcitation of isomeric state by EB mechanism}
First, we consider the deexcitation of the isomeric state of the $^{229}Th$ 
nucleus to its ground state using 
the generally accepted ordering of the levels shown in Fig. 16.  Fig. 6 gives 
the Feynman diagram describing the EB process.

The dominant nuclear transition is of M1 type.  The allowed E2 transition is 
damped by the ratio 
$({R_A \over \lambda_N})^2$, where $\lambda_N \gtrsim 2.5\cdot 10^{-5}$ cm is 
the wavelength of the nuclear radiation and $R_A$ is the nuclear radius.

In calculating the EB process probability, as before, we made a number of 
simplifications: we considered only the
direct diagram; here, as in many other cases (see, e.g., \cite{Kol90}) the 
exchange diagram is small.
Second, we used again the single-level approximation which is here well 
justified by the concrete analysis of the atomic
level structure \cite{Str91}.  The final goal is the calculation of the ratio 
$\eta$ for the probabilities $W_{\gamma}^{(3)}$ and
$W_{\gamma}^{(1)}$ introduced earlier (see Section 5, expressions (36), (37),
and (38) using the traditional 
method or (40) - (44) based on the ``discrete IC'' \cite{Zon90,Str91}).
In \cite{Str91} the initial electronic state was taken as $|i\rangle = 
6d_{3 \over 2}$, with energy 
$\epsilon_i$ = -4.2 eV, intermediate $|n\rangle = 7p_{1 \over 2}$, 
$\epsilon_n$ = -2.9 eV (values for $\epsilon_i$,
$\epsilon_n$ are theoretically estimated in \cite{Str91}).

Calculations \cite{Str91} have shown that $\eta$ is less than unity, i.e., the
direct nuclear deexcitation is the
dominant channel, if the nuclear transition energy difference $E_{is} = E_I - 
E_F$ is smaller than 
$\epsilon_n - \epsilon_i$ = 1.3 eV, leading to the expected lifetime of the 
isomer $T_{1 \over 2} \gtrsim 10$ d.

If the energy of the nuclear transition is at resonance with the energy of one
of the allowed atomic
transitions, $\Delta \equiv E_I - E_F - (\epsilon_n - \epsilon_i) \approx 0$ 
(e.g., $6d_{3 \over 2} \rightarrow
6d_{5 \over 2}$ or $7s_{1 \over 2} \rightarrow 6d_{3 \over 2}$), the EB 
probability increases sharply.  Of course,
ensuring the condition $\Delta = 0$ demands a much better knowledge of 
$E_{is}$ than is available today ((3.5 $\pm$ 1) eV
\cite{Hel94}).  For the range $2 \leq E_{is} \leq 5$ eV theoretical 
calculations give $T_{1 \over 2} \gtrsim 10$ min
\cite{Str91}, whereas the authors of \cite{Hel94} claim that $T_{1 \over 2} 
\gtrsim 45$ hr for $E_{is}$ = 3.5 eV
(M1 transition).  Taking into account the $\pm$1 eV uncertainty in the excited
state energy, $T_{1 \over 2}$ could be 
as long as $\approx$ 120 h or as short as $\approx$ 20 h.

However, one has to realize that the half-life time of an anomalously 
low-lying nuclear level ($E_{is} \approx$
several eV) is a very subtle quantity.  It is influenced by the interaction 
not only with the electronic shell,
but, due to the extremely small value of the isomeric state energy, it will be
affected also by the physical and
chemical properties of the sample containing the atoms of $^{229}Th$.  
Experimental and theoretical studies of such a phenomena may open new 
interesting directions into atomic and condensed matter physics.

We now turn to the question of the laser-assisted deexcitation of the isomeric
$^{229}Th$ state to its ground
state using the resonant, discrete IC with bound electrons, applying a laser 
field with the appropriate
frequency. As we saw above and in Section 5, this leads to a drastic 
acceleration of the nuclear isomeric
decay \cite{Str91,Kar92}.  Very recently, the probability for this process was
recalculated \cite{Typ96}.  As 
atomic levels, in contrast of \cite{Str91}, the states $7s_{1 \over 2}$ and 
$8s_{1 \over 2}$ were considered
with energies taken from \cite{Hua76} and \cite{Kar92}, respectively.  The 
energy difference of 3.713 eV between
these states is very close to the 3.5 eV nuclear decay energy.  This 
electronic transition (M1 excitation) requires
needs a photon energy of 0.1065 eV in a resonant two-photon process.  The 
experimental signal for the excitation
could be the photon emission during the decay of the excited 
$8s_{1 \over 2}$ state via the $7p_{1 \over 2}$ and $7p_{3 \over 2}$ states.

There is another possibility \cite{Kar92,Str91} based on the absorption of a 
single laser photon which excites the
$8s_{1 \over 2}$ state to the $8p_{1 \over 2}$ final state (first-order 
process in the laser interaction whereas
the first scheme corresponds to the second-order process).  In this case the 
necessary photon energy is 0.712 eV.

Because the probability increases in the first-order case only linearly with 
intensity, the second-order (in the laser
field) process was chosen in \cite{Typ96} since its probability has a 
quadratic increase and it is not small in 
absolute value.  Use of a resonant laser field is crucial since in the 
laser free case the deexcitation of the ${3 \over 2}^+$ 
nuclear state with excitation of the electron states given above has only a 
very small probability ($\sim 5\cdot 10^{-13}$) 
\cite{Typ96}.

Fig. 17 \cite{Typ96} shows the strong dependence of the deexcitation 
probability of the nuclear isomeric ${3 \over 2}^+$ state
of $^{229}Th$ for the $7s_{3 \over 2} \rightarrow 8s_{1 \over 2}$ M1 atomic 
transition on the
laser intensity $I$.  The laser photon energy is fixed to the energy of a 
resonant two-photon process with 
$E_{is}$ = 3.5 eV (solid line), 4.0 eV (long-dashed line), and 3.0 eV 
(short-dashed line) or for the resonant
four-photon process for an energy of 3.5 eV (dotted line).

We observe a strong increase with $I$, until the ionization thresholds of the 
electron states are reached
where the probability drops strongly.  As seen in Fig. 17, the absolute value 
of the deexcitation probability
depends on the laser photon energy.  One can also notice the quadratic 
dependence of the atomic shell excitation
probability on $I$ for the two-photon process and the fourth power of the 
intensity increase (with smaller probability).

In the paper \cite{Typ96} several examples
($^{161}Dy$, $^{189}Os$, $^{193}Ir$, $^{197}Au$,
$^{235}U$, $^{237}Np$) are considered from the point of view of the IEB 
mechanism leading to the excitation of nuclei.  The paper
considers the laser intensity up to $10^{24} {W \over cm^2}$.  We highly 
recommend this paper to the interested reader. 

\subsection{Pumping isomers by laser resonance}
The role of lasers is important also for the inverse process, i.e., the 
pumping of the ground state nuclei
$^{229}Th$ to its isomeric state.  As we stressed before, the $E_{is} = E_F - 
E_I$ is known today with poor
precision, which makes the direct application of the laser light to the 
nucleus in the ground
state ineffective.  The solution is to use the IEB mechanism which was 
discussed in Section 5 and which provides the
non-radiative excitation of the nucleus in an electron transition induced by 
the laser radiation field.  It is 
necessary to tune the laser light to the wavelengths of well known atomic
transitions.  In this case, even if 
there is a significant difference between the energies of the atomic 
($\epsilon_{in}$) and nuclear
($E_{IF}$) transitions involved, the excitation of the nucleus has a large 
enough probability to provide for an efficient pumping of the isomeric
state, which in turn opens the possibility to measure the energy and the 
lifetime of the low-lying nuclear isomeric state.

The (third order in the electromagnetic interaction) process is described by 
the Feynman diagrams in Figs. 14 and 15.
At the resonance ($\hbar\omega_L = \epsilon_n - \epsilon_i$) which provides 
more effective pumping, the 
(direct) diagram of Fig. 14 dominates and the single level approximation is 
reliable.  These features essentially simplify the calculations. 

Below, we follow the considerations given in the papers \cite{Tka92,Tk922}.  
For the wave functions of the initial ($i$),
intermediate ($n$), and final ($f$) shell electrons with energies 
$\epsilon_{i,n,f}$ and widths $\Gamma_{i,n,f}$ we take

\begin{equation}
\psi_{i,n,f}(\vec{R},t) = e^{-i(\epsilon_{i,n,f} - 
i{\Gamma_{i,n,f} \over 2})t}\psi_{i,n,f}(\vec{R}),
\end{equation}

\noindent
where the initial electronic shell $|i\rangle$ of the thorium atom is taken as 
$6d^27s^2$ with $\epsilon_i$ = -6.08 eV, and $\Gamma_i$ is taken to be zero.

Because the transition is of magnetic type (M1), we use the relativistic 
electronic wave functions $\psi(\vec{R})$
including the large $g(\vec{R})$ and small $f(\vec{R})$ component with the 
normalization condition
$$\int dR\cdot R^2(g^2(R) + f^2(R)) = 1.$$

The nuclear wave functions of the initial stationary state with energy $E_I$ 
and the final isomeric state
with energy $E_I$ and width $\Gamma_{is}$ are

\begin{eqnarray}
\psi_I(\vec{r},t) &=& e^{-iE_It}\psi_I(\vec{r}) \notag \\
 \\
\psi_F(\vec{r},t) &=& e^{-i(E_F - i{\Gamma_{is} \over 2})t}\psi_F(\vec{r}). 
\notag
\end{eqnarray}

A matrix element of the 3rd order process described by the direct diagram of 
Fig. 14 includes, in addition to the
wave functions for the electrons and the nucleus, the electromagnetic 
interactions of the photon with the electrons, 
$j^{\mu}(\vec{R})A_{\mu}(\vec{R})$, and nucleus $J^{\mu}(\vec{r})A_{\mu}
(\vec{r})$, respectively, where 

\begin{equation}
j_{if}^{\mu}(\vec{R}) = e\bar{\psi}_f(\vec{R})\gamma^{\mu}\psi_i(\vec{R})
\end{equation}

\noindent
is the electron electromagnetic current and 

\begin{equation}
J_{IF}^{\mu}(\vec{r}) = e\psi_F^+(\vec{r})\hat{J}^{\mu}\psi_I(\vec{r}),
\end{equation}

\noindent
where $\hat{J}^{\mu}$ is the four-vector of the nuclear electromagnetic 
current not specified here.
Furthermore, the matrix element includes the photon, $D_{\mu \nu}$, and the 
intermediate electron propagators.

The standard calculations include the expansion of the photon vector 
potentials and propagator in terms of 
multipoles (see, e.g., \cite{Akh65,Eiz70}). Furthermore, the matrix element 
of the nuclear current (49) is expressed in
terms of the reduced matrix elements \cite{Bla79}, taking into account the 
selection rules for the nuclear
transition, etc. Finally, one obtains the following expression for the
cross section $\sigma^{(3)}$ of the IEB process \cite{Tk922} at resonance with
the atomic $|i\rangle 
\rightarrow |n\rangle$ transition ($\hbar\omega_L = \epsilon_n - \epsilon_i$)
 ($\Gamma_i = 0$, $\Gamma_{is} \ll
\Gamma_n$):

\begin{multline}
\sigma^{(3)} = {\lambda_L^2 \over \pi}(1 + {\Gamma_f \over 2\Gamma_n})
{\Gamma_A^r(\omega_L; i \rightarrow n,
M(E)1) \over \Gamma_n} \\
\cdot{E_{int}^2(E_F - E_I; M1; n \rightarrow f, I \rightarrow F)
\over [\epsilon_n - \epsilon_f - (E_F - E_I)]^2 + \Gamma_n^2(1 + 
{\Gamma_f \over 2\Gamma_n})^2}
\end{multline}

\noindent
This expression is valid when the characteristic width of the laser line is 
comparable in magnitude with the atomic intermediate
state width $\Gamma_n$ : $\Delta\omega_L \approx \Gamma_n$.  Here, 
$\Gamma_A^r$ is the width of the atomic
radiative transition $i \rightarrow n$ of multipolarity $M(E)I$, and 
$E_{int}^2$ is the (energy)$^2$ of the
interaction of the electron transition current $j_{nf}^{\mu}(\vec{R})$ (48) 
and the nuclear current $J_{IF}^{\mu}(\vec{r})$ (49) averaged over the initial
and summed over the final states:

\begin{equation}
E_{int}^2 = \sum \begin{Sb}I  F \\n  f \end{Sb}|\int d^3r d^3R j_{nf}^{\mu}(R) 
 D_{\mu \nu}(\omega; \vec{R} - \vec{r})\cdot
J_{IF}^{\nu}(\vec{R})|^2,
\end{equation}

\noindent
where $D_{\mu \nu}(\omega; \vec{R} - \vec{r})$ is the Greens function of 
the photon in the coordinate-frequency
representation \cite{Ber82}

\begin{equation}
D_{\mu \nu}(\omega; \vec{R} - \vec{r}) = {-g_{\mu \nu}exp[i\omega |\vec{R} - 
\vec{r}|] \over |\vec{R} - \vec{r}|}.
\end{equation}

\noindent
With $|\omega_{IF}\cdot R| \ll 1$, (51) is reduced to the form 

\begin{equation}
E_{int}^2(\omega_{FI};M1) = {1 \over 4}\Gamma_A^r(\omega_{IF}; n \rightarrow f;
 M1) \cdot \Gamma_N^r(\omega_{IF};
I \rightarrow F, M1) (1 + \delta^2),
\end{equation}

\noindent
where $\Gamma_N^r$ is the probability for the radiative nuclear transition and

\begin{equation}
\delta = {Im[{\mathbf M}_{M1}(\omega_{IF})] \over Re[{\mathbf M}_{M1}
(\omega_{IF})]}
\end{equation}
 
\noindent
is the analog of the corresponding well-known quantity in the theory of IC, 
and 
\begin{equation}
{\mathbf M}_{M1}(\omega) = (\varkappa_n + \varkappa_f)\int_0^{\infty}dr\cdot 
r^2h_1^{(1)}(\omega r)[g_n(r)f_f(r) + f_n(r)g_f(r)],
\end{equation}

\noindent
where $h_1^{(1)}(x)$ is the spherical Hankel function and $\varkappa = 
(\ell - j)(2j + 1)$, and $\ell$ and $j$ are the orbital
and total angular momentum of the electron in the corresponding shell.  
The matrix element ${\mathbf M}_{M1}(\omega)$
is calculated numerically in \cite{Tka92,Tk922}.  Typical values for the 
important quantity $E_{int}^2$  
vary for different atoms in the range ($10^{-10} - 10^{-12}$) eV$^2$.  
In the already mentioned recent paper \cite{Typ96} this quantity was 
calculated for several nuclei on the basis
of the adiabatic description of the dressed electronic states in the laser 
field.  The values obtained there for $E_{int}^2$
coincide within one order of magnitude with the results of 
\cite{Tka92,Tk922}. However, in the
case of the popular $^{235}U$ the result is in drastic disagreement with the 
calculation of \cite{Tka92,Tk922}.

In the situations where the width $\Delta\omega_L$ of the laser is larger than
the widths of the atomic states 
$\Gamma_{n,f}$, it is necessary to carry out the integration of (50) obtained 
for monochromatic laser
photons over the shape of the laser line $g(\omega_L - \omega_0)$ with 
$\int g(\omega_L - \omega_0)d\omega_L = 1$.

At the resonance, $\omega_0 = \omega_{in}$, with the conditions 
$\Gamma_{n,f} \ll \hbar\Delta\omega_L$, and
$\Gamma_f, \Gamma_{is} \ll \Gamma_n \ll \epsilon_n - \epsilon_f - E_{FI}$, 
one has:

\small

\begin{equation}
\sigma_{res}^{(3)} \approx \lambda_L^2 {\Gamma_A^r(i \rightarrow n; 
M(E)1)E_{int}^2(\omega_{IF};M1;n \rightarrow
f, I \rightarrow F) \over [\epsilon_n - \epsilon_f - (E_F - E_I)]^2} \cdot
\begin{cases}
{1 \over \pi}& \text{~for $L$}\\ 
({\ln 2 \over \pi})^{1 \over 2}& \text{~for $G$}
\end{cases}, 
\end{equation}

\normalsize

\noindent
where L stands for Lorentzian and G for Gaussian line shapes. 

This expression, along with (50), (53), and (55), is the basis for the 
estimates below. 

\subsection{Pumping efficiency}
The arguments given in the previous subsection showed that the stimulation of 
the pumping of the isomeric state $^{229m}Th$ by the IEB
mechanism requires a careful study of
the structure and properties of the atomic levels of the thorium atom.  
However, information about the $^{229}Th$ atom is not complete.
In \cite{Cor68} only the energies of the levels are given, while in 
\cite{Gia74} energies, spins, and parities for some of the states are 
presented.

Two cases are interesting in the IEB mechanism for the thorium atom 
\cite{Tka92,Tk922}.\\
i) The atomic transition $|i\rangle \rightarrow |n\rangle$ has the 
multipolarity $M1$ (i.e., the same as the isomeric
nuclear transition $|I\rangle \rightarrow |F\rangle$) and the final state of 
the electron shell, $|f\rangle$,
coincides with the initial $|i\rangle$ one (``elastic'' IEB). The essential 
parameter $\Gamma_A^r$, defining
the cross section $\sigma^{(3)}$ of the IEB mechanism (see(50)) can be under 
these conditions $\sim(10^{-1} - 10^{-2})\Gamma_n$.
This situation is realized in the case, for example, 
$|n\rangle = 6d^37s(^5F_3)$, with $\epsilon_n$ =
-5.15 eV \cite{Rad86}.\\
ii) The $|i\rangle \rightarrow |n\rangle$ transition is of $E1$ or $M1$ type 
but 
the final state does not coincide with initial one (``inelastic'' IEB 
mechanism).
One possible realization is $|n\rangle = 6d7s^27p(^3F_2)$, $\epsilon_n$ = 
-4.74 eV and $|f\rangle = 5f6d7s^2(^3F_2)$,
$\epsilon_f$ = -5.06 eV \cite{Rad86}.
Under favorable conditions, for an atomic $E1$ transition $|i\rangle 
\rightarrow |n\rangle$ and small mismatch between
$E_F - E_I$ and the energy of one of the atomic transitions $|n\rangle 
\rightarrow |f\rangle$ ($\approx$ 0.1 eV)
the excitation cross-section due to the IEB mechanism may reach a value of the
order 
$\sigma_{res}^{(3)} \approx 10^{-20} - 10^{-21} cm^2$ for ordinary lasers with
${\Delta\omega_L \over \omega_L}
\approx 10^{-6}$.  At less favorable conditions, i.e., fixing the value for 
the mismatch $\tilde{\Delta} = E_{FI} - \epsilon_{nf}$
between the atomic and nuclear transitions in the energy denominators of Eqs. 
(50), (56) for 1 eV, for
$\omega_L$ = 1-5 eV and ${\Delta\omega_L \over \omega_L} \approx 10^{-6}$, and
taking for $E_{int}^2$ values between $10^{-10}$ and $10^{-12}$ it was found 
that $\sigma^{(3)} \approx 10^{-23}-10^{-25} cm^2$.

By moving ``up'' in the energies of the excited atomic states, one can find 
suitable atomic levels which should
be stimulated in the case $E_{is} >$ 2 eV.  Notice that the value for 
$\sigma^{(3)}$ found is a lower estimate.  Due
to the relatively high density of the excited atomic levels of $^{229}Th$, the
actual value of $\tilde{\Delta}$ may be much less than 1 eV.

For the elastic or inelastic $M1$ atomic transitions $|i\rangle \rightarrow 
|n\rangle$, the cross section
$\sigma^{(3)} \approx 10^{-25}-10^{-26} cm^2$ due to the small value of 
$\Gamma_A^r(M1) (\approx \alpha^2\Gamma_A^r(E1))$.

Defining the excitation efficiency $\xi$ as the ratio of the number of 
produced isomeric nuclei $^{229m}Th$
to the number of thorium atoms exposed to the laser light, we can write:

\begin{equation}
\xi = \sigma^{(3)}\tau\phi_L,
\end{equation}

\noindent
where $\tau$ is the irradiation time and $\phi_L$ is the flux density of the 
laser photons, $\phi_L =
{I \over \hbar\omega_L} = \rho_{\gamma}c$, and $\rho_{\gamma}$ is the laser 
photon density.  To lower the background
from the natural $\alpha$-decay of the thorium nuclei with the activity 
$A_{Th}(\alpha) = 3.2\cdot 10^{-12} N(Th)$
decays per second, where $N(Th)$ is the number of thorium nuclei, let us 
estimate the induced activity for a very
small amount of thorium atoms taking a sample of mass $10^{-8}$ g evaporated 
as a layer of thickness $10\AA (N(Th) = 10^{13})$ onto a backing of diameter 
1 mm.  Under these conditions a fairly small 
absorption of the photons emitted in the decay of the $^{229m}Th$ isomers in 
the sample is guaranteed.

Irradiation of this target with a laser beam of power 100-200 mW focused onto 
$\approx$ 1 mm for the $\tau = 10^2-10^3$s
gives the efficiency $\xi \approx$ 0.01-1 with $\sigma^{(3)} = 
10^{-23}-10^{-24} cm^2$.  As a result, the 
$\gamma$-activity of the isomer $^{229m}Th$ will be $10^5-10^7$ Bq (for the 
$T_{1 \over 2}^{is} \approx$ 10 d).
Thus, using a tunable laser one can stimulate a suitable atomic state with 
energy greater than the energy of the
isomeric transition $E_F - E_I$, and the nuclear excitation will take place 
via the decay channels of this atomic state.

In conclusion, the example of the $^{229}Th$ nuclei shows that the optical 
laser pumping of low-lying isomeric nuclear
states via the IEB mechanism is a realistic and relatively simple tool to 
measure and study lifetimes and energy levels. 

\subsection{Modified IEB Mechanism in the study of $^{229m}Th$}
There exists a modification to the use of the IEB mechanism for the excitation
and study of the properties of the
isomeric state of the thorium nucleus \cite{Kal94}.  Below, we follow the 
paper \cite{Kal94}.  It is connected
with the study of the photon energy which emerges during the nuclear 
excitation.  Schematically, it looks as follows
(Fig. 18): A laser beam excites the atomic electron to a state of angular 
momentum defined by the selection rules in
the subsequent deexcitation of the shell.  The electronic deexcitation leads 
to the excitation of the nucleus and,
simultaneously, to the photon emission.

In the following, we assume that the electron has already reached the excited 
state $|n\rangle$.  The nuclear
quadrapole transition ${5 \over 2}^+ \rightarrow {3 \over 2}^+$ (or ${3 \over 
2}^+ \rightarrow {5 \over 2}^+$,
if the ground nuclear state of $^{229}Th$ has angular momentum ${3 \over 2}^+$
and not ${5 \over 2}^+$ as
it is usually claimed), with photon emission in the dipole regime, imposes the 
condition $j_i + j_n \geq 3$.

Thus, a new line appears in the optical spectrum of the laser excited 
$^{229}Th$
spectrum, its frequency is defined
by the relation

\begin{equation}
\hbar\omega =\hbar\omega_{ni} - \hbar\omega_{FI}.
\end{equation}

\noindent
That opens the possibility of the \\
i) precise determination of the energy of the $^{229m}Th$ isomeric nucleus by 
measuring the photon frequency $\omega$;\\
ii) solution of the uncertainty concerning the question of the angular momenta
of the nuclear ground and isomeric
states (${5 \over 2}^+$(or ${3 \over 2}^+$) and ${3 \over 2}^+$ (or ${5 \over 
2}^+$), corresponding to the
sign of the difference $\omega_F - \omega_I \lessgtr 0$).  Furthermore, once 
the magnitude of the emitted photon
frequency $\omega$ is known, the process can be accelerated by the resonant 
application of the laser with 
frequency $\omega_L = \omega$.  The simplified calculations are very similar 
to the calculations of Sections 4 and 5
for the IC and EB processes. The role of the initial electronic state 
$|i\rangle$ plays here the laser-excited,
dressed electronic state $|n\rangle$.  Again, emission of the photon frequency
$\omega$ is treated quantum-mechanically.

The nuclear transition is a quadruple type, L=2.  The initial, excited atomic 
state is taken as:
$$i: |n\rangle = 8p7s^2, n_i = 8, l_i = 1.$$
We take the final electronic state $|f\rangle$ as
$$f: |f\rangle = 6d^27s^2$$ with $n_f = 6, l_f = 2$
satisfying the above mentioned condition about the angular momenta involved.

Since the initial electronic state in our scheme is the excited state, we have
to take into account the energy
distribution of this state around its central energy $\epsilon_{i0}$ with width
 $\gamma_i$ (line shape)

\begin{equation}
\rho(\epsilon_i) = {\gamma_i \over 2\pi}{1 \over (\epsilon_i - \epsilon_{i0})^2
 +{\gamma_i^2 \over 4}}
\end{equation}

\noindent
as a weight factor in the integration (averaging) over the initial electronic 
state.  In the resonance case we have
$\epsilon_{i0} = \epsilon_f + E_F - E_I + \hbar\omega_L$.  The integration 
gives a factor ${2 \over \pi}\gamma_i^{-1}$.  The laser induced IEB process at 
$\omega_L = \omega$ can be written as

\begin{equation}
W_{fi}^L = {1 \over 2\pi}W_{fi}\phi\lambda_L^2{1 \over \gamma_i},
\end{equation}

\noindent
where $\phi$ is the laser flux ${Nc \over V}(= {I \over \hbar\omega_L})$ 
($V$-volume), and $W_{fi}$ is the 
transition probability (per unit time) of the spontaneous photon emission due 
to the IEB:

\begin{equation}
W_{fi} = CW_{fi}^{\circ}|\langle F||\hat{Q}_L||I\rangle|^2
|I_{L,n_fl_f}^{n_il_i}|^2,
\end{equation}

\noindent
where

\begin{equation}
W_{fi}^{\circ} = ({r_0 \over a_B})^{2L} {e^6 \over m^2c^3a_B^4} {8\pi^2(L + 1) 
\over (2L + 1)(2l_i + 1)}
{C_1(L) \over \hbar\omega}
\end{equation}

\noindent
and $|\langle F||\hat{Q}_L||I\rangle |^2$ is the squared reduced matrix 
element of the multipole moment of order $L$ between the nuclear states 
$|I\rangle$ and $|F\rangle$, and is expected to have a value between $10^{-2}$
and $10^2$ (see, \cite{Eis70}).

In (61), $I_{L,n_fl_f}^{n_il_i} = a_B^{L + 2}\int r^{1 - L}R_{n_il_i}
(Z_{eff}^i,r)R_{n_fl_f}(Z_{eff}^f,r)dr$ (cf. (37)) which for our case 
($n_i = 8$, $l_i = 1$, $L = 2$, $n_f = 6$, $l_f = 2$,
$Z_{eff}^i = 2.17$, $Z_{eff}^f = 4.65$) gives $|I_{2,62}^{81}|^2 = 4.49 \cdot 
10^{-4}$.  $C$, $C_1(L)$ are constants expressed in terms of $3j$-symbols.

For $W_{fi}^{\circ}$ we obtain:

\begin{equation}
W_{fi}^{\circ} = 8.7\cdot 10^{-5}{1 \over \hbar\omega}, \text{in} \quad s^{-1}
\end{equation}

\noindent
and 

\begin{equation}
W_{fi} = 1.7 \cdot 10^{-8}|\langle F||\hat{Q}_L||I\rangle|^2{1 \over 
\hbar\omega}, \text{in} \quad s^{-1}.
\end{equation}

The ground state of $^{229}Th$ has a half-life time $T_{1 \over 2} \approx 2.3
\cdot 10^{11} s$ \cite{Led78}.
A sample of N $^{229}Th$ atoms has a radioactivity $A = 8.1 \cdot 10^{-23} N$ 
(in units of Ci).  Thus, for a sample
with $N = 10^{18}$ atoms we have $A = 8.1 \cdot 10^{-5}$ Ci.  If we suppose 
that 1\% of the atoms can be populated
in the desired electronic initial state by the laser field, then the photon 
emission rate $R$ (in $s^{-1}$) induced by the IEB process is 

\begin{equation}
R = W_{fi} \cdot 10^{16} = 1.7 \cdot 10^8|\langle F||\hat{Q}_L||I\rangle|^2 {1 
\over \hbar\omega}
\end{equation}

\noindent
which is larger than $10^{6} s^{-1}$.  It is a lower bound since one can 
expect that the state $|F\rangle$ has
a much smaller half-life time than the state $|I\rangle$, thus the activity of
the sample will be much larger.

If we compare these estimates with estimates discussed at the end of 
Subsection 6.4 for a somewhat different
process, where $N$ was $10^{13}$ and the obtained activity was $10^5-10^7$ Bq 
for an irradiation time $\sim 10^3$ s, 
we obtain here an induced rate $R \sim 10^8-10^{10} s^{-1}$ or, for 
$N = 10^{18}$, as used above, $R \approx 10^{13}-10^{15} s^{-1}$.

These numbers look encouraging and show that the laser-driven IEB mechanism 
can offer a reliable method for
determining accurately the energy difference and other properties such as 
lifetime and angular momentum of ground and 
isomeric states, etc., and controlling a radioactive decay rate.

One can go further and use the small energy difference of the ${3 \over 2}^+$ 
and ${5 \over 2}^+$ levels of the
$^{229}Th$ nucleus to modify the $\alpha$-decay rate of the nucleus in a laser
driven resonant process where
these two nuclear levels are mixed by the magnetic field of the laser field 
\cite{PKA}.  As a result, the $\alpha$-decay
rate of the $^{229}Th$ nucleus to one of the levels of the daughter nucleus 
can be written at the resonance as

\begin{equation}
R = R_{5 \over 2} + \nu R_{3 \over 2}
\end{equation}

\noindent
where $R_{{5 \over 2}({3 \over 2})}$ is the $\alpha$-decay rate for the 
${5 \over 2}({3 \over 2})$ state in the 
laser free case,

\begin{equation}
\nu = {\pi^2e^2I \over M^2c^3\Delta\omega_L}\beta\tau,
\end{equation}

\noindent
where $M$ is the nucleon mass, $\Delta\omega_L$ is the band width of the 
laser, $\tau$ is 
the irradiation time and $\beta$ is the reduced M1
transition matrix element in units of the Bohr magneton.  Here. $\beta$ is of 
order $10^{-4}$ (see, e.g. \cite{Eiz70}).
Assuming that the lifetime of the $^{229m}Th({3 \over 2}^+)$ is much shorter 
than that of the ground state 
$({5 \over 2}^+)$ and taking $\Delta\omega \approx 5.7\cdot 10^6 s^{-1}$, 
$\beta \sim 10^{-6}$, $\tau = 1 h$,
$I = 10^3 {W \over cm^2}$, one obtains $\nu = 0.19$ \cite{PKA}.  This last 
example\footnote{Such kind of effects,
of course, necessitate further detailed and accurate investigation.  
Particularly, in the considered example
the screening of the magnetic field by atomic shells seems essential for the 
level mixing.} shows once more the
potency of the laser assisted EB and IEB mechanisms for modifying and 
controlling nuclear radioactivity by the influence of light - a topic which
goes back to Einstein, as shown in the epigraph to the present paper.

\section{Summary and conclusions}
In the present paper we have tried to review several applications of optical 
and UV-lasers for studying low-energy properties of nuclei.

The most effective tool for these applications is the use of the electronic 
shells of atoms as mediators 
between the laser field of appropriate frequency and the nuclei of interest. 

All effects considered above are united by the realization of two unique 
advantages of lasers (besides other
important properties like monochromaticity, polarization, etc., shared with
other sources of electromagnetic radiation):\\
i) High intensity of the laser beam makes the multiphoton absorption 
(emission) by atomic shells possible, thus 
providing the effective elimination of the mismatch between the energy 
differences of atomic and nuclear
transitions and, furthermore, leading to resonantly enhanced effects.  This 
possibility of using multiples of the laser
frequency $\omega_L$ to provide the transition energy is the leading principle
for the use of intense
laser beams for the study of low-energy nuclear processes and properties;\\
ii) The dipole character of the interaction of the laser radiation with the 
atomic shells provides a reduction 
of the (usually high) multipolarity of the nuclear electromagnetic radiation, 
thus effectively enhancing
the transition probability (the laser light transfers not only energy, but 
also angular momentum).

It was shown that, for intense laser beam ($I > 10^{19} {W \over cm^2}$), the 
so called ``anti-Stokes''
transition, although not so probable as it was expected before, provides the 
effective mechanism for the
deexcitation of nuclear levels located close enough to the nuclear ground 
state.  
This mechanism can be used
to study specific nuclear levels and their transitions 
unaccessible otherwise.  It was shown
that the laser-assisted (induced) internal conversion (IC) is effective only 
for the nuclei where IC is 
forbidden by energy conservation in the laser-free cases.
Close and below the threshold for IC the role of optical and UV-lasers is very
important and leads, in principle, to
observable laser induced IC coefficients.  Unfortunately, for very intense 
lasers the hindering
effects of pondermotive forces have not been taken into account yet.  Thus, 
the theoretical study of the IC processes 
at very high $I$ is important.  Generally, this case demands the elimination 
of several simplifications that we used for
the study of this process. Here, time-consuming numerical calculations are
required.

We have studied the role of the EB and IEB mechanisms in processes of resonant
deexcitation and excitation
of isomeric nuclei with emphasis on measuring their energies and lifetimes.  
Special 
attention was paid to the study of the low-lying isomeric states of the 
nucleus $^{229m}Th$ where the use
of optical and UV-lasers seems the most promising.

The existence of anomalously low-lying excited nuclear levels (as $^{229m}Th$)
opens the real possibility to control
and modify nuclear processes by optical and UV-lasers.  Furthermore, the great
sensitivity of these low-lying nuclear levels to the physical and chemical 
environment could
provide valuable information for atomic and condensed matter physics, optics, 
plasma physics, lasers, etc.
Almost all processes considered here were treated using many simplified 
approximations, thus one cannot expect
to achieve precise quantitative conclusions.  The present values for the 
transition probabilities of many processes are rather 
uncertain and may change considerably if more exact calculations of the 
electronic transition matrix elements become available in the future.

Of particular concern is the assumption of independent one-electron levels 
used in all cases above, which limits
the predictive power of the calculations considerably.  Due to many body 
effects, any perturbation of neighboring electronic shells by the intense 
laser radiation
will have an influence on the binding energy and wave function of each 
electron.  
However, we believe that the essential aspects of the laser-assisted and 
laser-induced nuclear processes
studied here will turn out to be correct and will provide useful directions 
for future studies.

>From the experimental point of view, it is necessary to 
search for the most favorable low multipolarity transitions between inner 
electron states to achieve the strong nucleus-electron shell coupling.

In present (and future) applications, especially when the fulfillment of the 
resonance conditions by laser fine tuning
becomes experimentally feasible, the knowledge of the precise energies of the 
nuclear and atomic states involved is of fundamental importance.

\section*{Acknowledgments}
It is a pleasure to thank Werner Tornow, who attracted the author's attention 
to the subject, for discussions, support, and interest in the present work.  I
am grateful to Vladimir Litvinenko for explanations and discussions
of the operation, present status and future of the Duke University FEL 
Facility, to Berndt M\"uller for discussion
of theoretical topics, interest and encouragement, and to Karl Straub, who 
attracted my attention to the role
of laser radiation in the orbital electron capture process by nuclei, and for 
his general interest in the present work.
I also thank Holly Pulis for helping to prepare this manuscript.
This work was supported in part by the US Department of Energy, Office of 
High Energy and Nuclear Physics, under
grant No. DEFG05-91ER40619.

\section*{\underline{Appendix A}: Parameters characterizing the interaction of
the laser radiation with electron
(dimensional arguments)}
\addcontentsline{toc}{section}{\numberline{}Appendix A: Parameters 
characterizing the interaction of the laser 
radiation with electron (dimensional arguments)}
Here we use dimensional arguments to show the role of the most important 
parameter of the intense
laser field, the intensity, in interaction processes with electrons 
\cite{Ebe69}.

The intensity $I = \hbar\omega_L\rho c$ ($\rho$ is the number density of laser
photons) must, of course, play the
essential, nontrivial role in such physical quantities as transition 
probability, cross section, etc. However, its
inclusion should not change the dimensionality of the expressions for such 
quantities.  This means that the 
density $\rho$ must always enter in the combination $\rho L^3$ where $L$ is 
some length characteristic to the
effect under consideration.  There are several lengths 
(for free electron and monochromatic radiation):\\
-classical radius of the electron
$$r_0 = {e^2 \over mc^2} = 2.8 \cdot 10^{-13} cm$$
-wavelength of the laser radiation
$$\lambda_L (\approx 10^{-4}-10^{-5} cm), \text{or} \lambdabar_L \equiv
 {\lambda_L \over 2\pi}$$
-(reduced) electron Compton wavelength 
$$\lambdabar_c = {\hbar \over mc} = 3.86 \cdot 10^{-11} cm.$$

The first two of them are classical quantities, $\lambdabar_c$ is of quantum 
nature.
Following \cite{Ebe69}, it is illuminating to separate the most interesting 
dimensionless combinations of
$\rho$ (which is a quantum quantity since it measures the numbers of photons 
($\rho = {I \over \hbar\omega_L c}$))
and these three lengths into classical and quantum parameters:\\
We have (retaining for clarity $\hbar$ and c):\\
Classical dimensionless parameters:

\begin{align}
&C_1: \quad \rho\lambdabar_c r_0^2 = {Ir_0^2 \over \omega_L mc^2} \notag \\
&C_2: \quad \rho\lambdabar_c r_0\lambda_L = {Ir_0\lambdabar_L^2 \over mc^3} = 
10^{-10}\lambda_L^2I \notag \\
&C_3: \quad \rho\lambdabar_c \lambda_L^2 = {I\lambdabar_L^2 \over \omega_L 
mc^2} \notag \\
\intertext{Quantum dimensionless parameters:} 
&Q_1: \quad \rho\lambdabar_L^3 = {I\lambdabar_L^3 \over \hbar\omega_L c} 
\notag \\
&Q_2: \quad \rho\lambdabar_L^2r_0 = {I\lambdabar_L^2r_o \over c}{1 \over 
\hbar\omega_L} \notag
\end{align}

The parameter $C_1$ is extremely small. The parameter $C_2$ has an interesting
physical meaning, defining the
ratio of the electromagnetic energy $r_0\lambdabar^2{I \over c}$ of the wave 
with volume $r_0\lambdabar^2$
passing by (in ``contact'' with) the electron of size $r_0$, to the electron 
rest energy.  We see that the 
relativistic effects described by $C_2 = 10^{-10}\lambda_L^2I$ are small for 
the $\lambda_L$ of interest here
($\lambda_L \sim 10^{-4}-10^{-5} cm$) unless $I > 10^{10} {W \over \lambda^2} 
\approx 10^{20} {W \over cm^2}$.

The parameter $C_3$ has the same meaning as $C_2$ but with respect to a much 
larger volume $\lambdabar^3$.  It seems to play no role since it is difficult 
to imagine an electron to be probed in the interaction
region as large as $\lambdabar^3$.  The interesting quantum parameter is $Q_2$
which measures the ratio
of the electromagnetic energy ${r_0\lambdabar^2I \over c}$ to the photon 
energy $\hbar\omega_L$, i.e.,
defines the number of the laser photons in the interaction region 
$r_0\lambdabar^2$.  In other words, just
this parameter describes the nonlinear, multiphoton interactions with free 
electron.  We note that these kinds of
effects were important in several processes described in the text, the only 
difference is that such multiphoton
processes referred to bound electrons.  The parameter $Q_1$ has the same
meaning as $Q_2$, again with respect to the larger volume $\lambdabar^3$.  

The above arguments were applied to the interaction of laser radiation with 
free electrons.  

In some sense, the dimensional arguments are useful for atomic systems even 
under the influence of very intense lasers with an
electric field strength $E_0$ of the order of $(1-5){e \over a_B^2}$ (i.e., 
$I = (7 \cdot 10^{16}-2 \cdot 10^{18}) {W \over cm^2}$).
With laser photon energy $\hbar\omega_L$ = 5 eV the amplitude of the motion of
the electrons is larger than
${eE_0 \over m\omega_L^2} \sim 30a_B$.

These arguments were the basis for the authors of \cite{Bar88,Ber91} to 
construct an approximate theory of  
nuclear excitations due to the motion of the shell electrons in the very 
intense field of the optical and UV
lasers when a laser driven highly localized current of collectively 
oscillating electrons induces the nuclear
excitation.  Each electron in this current moves on its own classical 
trajectories (see also \cite{Sol88,Rin88,Rin87}).
Such a model gives the upper limit for the effect of nuclear excitations 
\cite{Bar88}.

In conclusion, it is worthwhile to remark that for the atomic electrons there 
are additional characteristic lengths
such as the Bohr radius $a_B$ or ${a_B \over Z}$, so one could expect that the
corresponding important parameters
$C_2$ and $Q_2$ will be much larger (see also Section 4 which is devoted to 
the laser assisted and induced
IC processes).  
Thus, the interaction of laser radiation of optical and UV range with a simple
atomic systems would
offer a very promising way to study the nonlinear effects intrinsic to the 
intense laser field.  This issue
was illustrated in the text from the point of view of low-energy nuclear 
physics.

\section*{Appendix B: Two representations of the interaction of radiation with
matter}
\addcontentsline{toc}{section}{\numberline{}Appendix B: Two representations of
the interaction of the radiation
with matter} 
As it was pointed out in the Introduction (footnote 1), the question of  
relating the dipole form (1) as the
Hamiltonian of the matter-radiation field interaction to the general gauge 
invariant (non-relativistic)
Hamiltonian $H = {1 \over 2m}(\hat{\vec{p}} - e\vec{A})^2 (c = 1)$ which gives 
the interaction term

\begin{equation}
-{e \over m}\vec{\hat{p}}\cdot \vec{A} + {e^2 \over 2m}\vec{A}^2 \tag{B1}
\end{equation}

has a long and controversial history (see \cite{WEL}).  At first glance, the 
problem is simply solved by  
applying to the wave function $\psi$ in the Shr\"odinger equation 
${i{\partial \psi} \over \partial t} = H\psi$ the unitary transformation

\begin{equation}
\psi(\vec{r},t) = exp\{i\vec{A}(t)\cdot\vec{r}\}\phi(\vec{r},t) \tag{B2}
\end{equation}

(in the dipole approximation the vector potential $\vec{A}$ does not depend 
on $\vec{r}$).  We obtain for $\phi(\vec{r},t)$ the equation

\begin{equation}
i{\partial \phi \over \partial t} = [{1 \over 2m}\hat{p}^2 + \vec{r}\cdot
\vec{\dot A}(t)]\phi \tag{B3}
\end{equation}

\noindent
which leads, using $\vec{E} = -\vec{\dot{A}}(t)$, to the G\"oppert-Mayer form 
(1).  Obviously, if
$\vec{A}$ depends on $\vec{r}$, i.e., we avoid the dipole approximation, this 
correspondence is not valid.
This equivalence is well known.  But this is not the whole story.
First of all, one needs to be careful if in actual studies (practically 
unavoidable) 
approximations (besides the dipole one) have to be made.  Eqs. (B3) and (1) 
are equivalent in the dipole 
approximation if the wave functions $\psi$ and $\phi$ are exact solutions of 
the corresponding Schr\"odinger 
equations.  If approximations are permitted, the equivalence may be destroyed.
This situation leads, for example,
to the conclusion \cite{Del85} that when studying model systems with a finite 
number of levels, one must choose
the dipole interaction in the form (1), rather than in the form (B3).

However, even if the $\psi$ and $\phi$ are the exact solutions of the 
Schr\"odinger equation, one needs to be 
cautious when the problem includes damping which, in the case, e.g., of the 
two-level system (described by a  2$\times$ 2
matrix) has off-diagonal elements for the $\vec{p}\cdot\vec{A}$ interaction. 
It makes the dipole form 
$-e\vec{r}\cdot\vec{E}$ more convenient, although, if one takes into account 
the proper change of the damping
matrix under the transformation (B2), they are equivalent.  

Since this point is not emphasized in textbooks we give here the derivation 
\cite{WEL} of the above statement
for the two-level system studied in Section 3.  Consider the Bethe-Lamb 
equations (7) for two states
$|a\rangle$ and $|b\rangle$ with decay constants $\gamma_{a(b)}$, $E_a - E_b =
\hbar\omega_a - \hbar\omega_b =
\hbar\omega$, under influence of the radiation field $\vec{E}(t) = \vec{E}_0
\sin \nu t$ with Hamiltonian in
the dipole form: $H = -e\vec{E}\cdot\vec{r}$, we have:

\begin{gather}
\dot{a} = -{\gamma_a \over 2}a + {i \over \hbar}e\vec{E}_0\cdot\vec{r}_{ab}
\sin\nu t e^{i\omega t}b \notag \\
 \tag{B4} \\ 
\dot{b} = -{\gamma_b \over 2}b + {i \over \hbar}e\vec{E}_0\cdot\vec{r}_{ba}
\sin\nu t e^{i\omega t}a \notag
\end{gather}

\noindent
($\vec{r}_{ab} = \langle a|\vec{r}|b\rangle$).  These equations are derivable 
from the Schr\"odinger equation

\begin{equation}
i\hbar{\partial \psi \over \partial t} = [H_0 - {1 \over 2}i\hbar\Gamma - 
e\vec{E}(t)\cdot\vec{r}]\psi \tag{B5}
\end{equation}

\noindent
where $H_0$ is the unperturbed Hamiltonian $H_0 = {\vec{p} \over 2m} + 
U(\vec{r})$ with eigenstates $\psi_a$ and
$\phi_b$:

\begin{equation}
H_0\psi_a = \hbar\omega_a\psi_a, H_0\psi_b = \hbar\omega_b\psi_b \tag{B6}
\end{equation}

\noindent
and $\Gamma$ is the Weisskopf-Wigner type decay operator with eigenvalues 
$\gamma_a$, $\gamma_b$:

\begin{equation}
\Gamma\psi_{a(b)} = \gamma_{a(b)}\psi_{a(b)} \tag{B7}
\end{equation}

Obviously, if we insert the wave function of our two-level system

\begin{equation}
\psi(\vec{r},t) = a(t)e^{-i\omega_a t}\psi_a(\vec{r}) + b(t)
e^{-i\omega_b t}\psi_b(\vec{r}) \tag{B8}
\end{equation}

\noindent
into (B5) we obtain the Bethe-Lamb equations (B4).

Now, to replace the $\vec{E}\cdot\vec{r}$ term in (B5) by the $\vec{p}
\cdot\vec{A}$-type interaction (B1) and we make the transformation (B2):

\begin{equation}
\psi(\vec{r},t) = T^+(\vec{r},t)\phi(r,t), \tag{B9}
\end{equation}

\noindent
with $T(\vec{r},t) = e^{{ie \over \hbar}\vec{A}(t)\cdot\vec{r}}$ (we retain 
here $\hbar$), thus obtaining the equation for $\phi$:

\begin{equation}
i\hbar{\partial \phi \over \partial t} = \{H_0 - {i\hbar \over 2}
\Gamma'(r,t) - {e \over m}\vec{p}\cdot\vec{A}
+ {e^2 \over 2m}\vec{A}^2\}\phi. \tag{B10}
\end{equation}

\noindent
Thus, we arrive at the important result that in the equivalence stated above 
between the two forms of interactions (in
dipole approximation) it is necessary to treat the transformed damping 
operator $\Gamma'$

\begin{equation}
\Gamma' = T\Gamma T^+ \tag{B11}
\end{equation}

\noindent
In contrast to equations (B4), $\Gamma'$ depends on $\vec{r}$ and $t$ and is 
not diagonal with respect to the
eigenstates $\psi_{a(b)}$ of $H_0$.  Indeed, we have:

\begin{equation}
\Gamma = \begin{pmatrix} \gamma_a & 0 \\ 0 & \gamma_b\end{pmatrix} \tag{B12}
\end{equation}

\begin{equation}
\Gamma' = \begin{pmatrix} \gamma_a\cos^2\phi + \gamma_b\sin^2\phi & 
-i(\gamma_a - \gamma_b)\sin\phi\cos\phi \\
i(\gamma_a - \gamma_b)\sin\phi\cos\phi & \gamma_a\sin^2\phi + 
\gamma_b\cos^2\phi\end{pmatrix}, \tag{B13}
\end{equation}

\noindent
where $\phi = e\vec{A}(t)\cdot{\vec{r}_{ba} \over \hbar}$.

These results show that it is wrong to simply replace the 
$\vec{E}\cdot\vec{r}$ interaction by the
$\vec{p}\cdot\vec{A}$ form; the damping matrix has to be changed in addition. 
The off-diagonal elements of the
new damping matrix $\Gamma'$ contribute both to the coupling of the $\psi_a$ 
and $\psi_b$ states.  Finally, it is 
instructive to write the Bethe-Lamb type equations for the amplitudes $\alpha$
and $\beta$ in the expansion $\phi$ in terms of $\psi_a$ and $\psi_b$:

\begin{gather}
\dot{\alpha} = -({1 \over 2}\Gamma'_{aa}(t) + {i \over \hbar}{e^2 \over 2m}
\vec{A}^2(t))\alpha - ({1 \over 2}\Gamma'_{ab}(t) - {i \over \hbar}{e \over m}
\vec{p}_{ab}\cdot\vec{A}(t))e^{i\omega t}\beta \notag \\
 \tag{B14} \\
\dot{\beta} = -({1 \over 2}\Gamma'_{bb}(t) + {i \over \hbar}{e^2 \over 2m}
\vec{A}^2(t))\beta - ({1 \over 2}\Gamma'_{ba}(t) - {i \over \hbar}{e \over m}
\vec{p}_{ba}\cdot\vec{A}(t))e^{i\omega t}\alpha,  \notag
\end{gather}

\noindent
where $\vec{p}_{ab} = \langle a|\vec{p}|b\rangle$.

We note that in the calculations of the probability amplitudes in the two 
representations the initial conditions
for the equations of motion (B14) are in general different from the initial 
conditions of the Bethe-Lamb equations (B4) \cite{WEL}.

A last remark concerns a practical issue. For higher order processes it 
becomes more and more cumbersome to solve the $\vec{p}\cdot\vec{A}$
equations (B14) due to the transformed damping matrix elements. Therefore, for 
practical calculations even for a two-level atom the use of the Bethe-Lamb 
equations (B4) is more convenient.
\newpage
\section*{Notation Index}
\addcontentsline{toc}{section}{\numberline{}Notation Index}
$\lambda_L$ - laser radiation wave length measured in nm or cm\\
$\lambdabar_L = {\lambda_L \over 2\pi}$\\
$\Delta\lambda_L$ - laser radiation linewidth\\
$\hbar\omega_L$ - energy of laser photon measured in eV\\
$\lambda_{ab} \equiv {2\pi\hbar c \over E_a - E_b}$\\
$\omega_L$ - frequency of laser photon\\
$k_L$ - wave number of laser photon\\
$R_A$ - nuclear radius\\
$\Delta E_A$ - nuclear level spacing\\
$H$ - Hamiltonian\\
$\alpha = {e^2 \over \hbar c}$ - fine structure constant\\
$\vec{R}$ - radius-vector of atomic electron\\
$\vec{r}_p$ - radius-vector of the p-th nuclear proton\\
$\vec{E}_L$ - electric field of laser radiation\\
$\vec{E}_0$ - amplitude of laser electric field\\
$I$ - laser beam intensity measured in units of ${W \over cm^2}$\\
$\phi$ - photon flux\\
$\mathbf{M}$ - matrix element\\
$\sigma$ - cross section\\
$\gamma$ - level-decay rate\\
$\Gamma = \hbar\gamma$ - level width\\
$N$ - number of laser photons\\
$|I\rangle$ - initial nuclear state\\
$|F\rangle$ - final nuclear state\\
$|N\rangle$ - intermediate nuclear state\\
$E_I$ - energy of initial nuclear state\\
$E_F$ - energy of final nuclear state\\
$E_N$ - energy of intermediate nuclear state\\
$|i\rangle, |n\rangle, |f\rangle$ - initial, intermediate, and final 
atomic electronic state, respectively\\
$\epsilon_i, \epsilon_n, \epsilon_f$ - energy of initial, intermediate, 
and final atomic electrons, respectively\\
$E_{IF} \equiv E_I - E_F$\\
$\epsilon_{if} \equiv \epsilon_i - \epsilon_f$\\
$E_{is}$ - energy of nuclear isomeric states\\
EB - electron bridge\\
IEB - inverse electron bridge\\
IC - internal conversion\\
$\Delta E$ - nuclear level energy difference\\
$\Delta\epsilon$ - atomic electron level energy difference\\
$m(M)$ - electron (proton) mass\\
$Z$ - atomic number\\
$Z_{eff}$ - effective atomic number\\
$E_B$ - electron binding energy in atomic shell, sometimes we use $|E_B| 
\equiv B$\\
$c$ - speed of light\\
$\vec{A}$ - vector potential of electromagnetic field\\
$\rho(\epsilon_f)$ - density of final electron states\\
$L$ - multipolarity of nuclear radiation\\
$l$ - orbital angular momentum\\
$m$ - magnetic quantum number\\
$n$ - principal quantum number\\
$\bar{P}$ - laser average power\\
$\bar{J}$ - laser average current\\
$\hat{P}$ - laser peak power\\
$\rho$ - number density of laser photons\\
$r_0 = {e^2 \over mc^2} = 2.8 \cdot 10^{-13}$ - classical electron radius\\
$\lambdabar_c = {\hbar \over mc} = 3.86 \cdot 10^{-11} cm$ - electron 
Compton wavelength\\
$a_B = {\hbar^2 \over me^2} = 5.3 \cdot 10^{-9} cm$ - Bohr radius\\
$R_y = {e^2 \over 2a_B}$ - Rydberg potential\\
$\vec{p} = \hbar\vec{k}$ - electron momentum\\
$\vec{k}$ - wave vector\\
$E(M)L$ - electric (magnetic) transition with multipolarity $L$\\
$J_{I(F)} (J_{i(f)})$ - nuclear (atomic) angular momentum for initial 
(final) states\\
$m_{I(F)}(m_{i(f)})$ - nuclear (atomic) magnetic quantum numbers for 
initial (final) states\\
$\Delta$ - energy mismatch; detuning\\
$T$ - time interval, temperature\\
$\alpha_{IC}$ - coefficient of internal comversion\\
$\alpha_d$ - coefficient of IC for discrete final electron state\\
$n, j, \lambda$ - quantum numbers of hydrogen-type solution\\
$\vec{e}_1, \vec{e}_2, \vec{e}_3$ - unit vectors defining the photon 
polarization

\newpage
\addcontentsline{toc}{section}{\numberline{}References}

\newpage
\section*{Figures}

\vspace{3in}

\begin{center}
\parbox{\textwidth}{Fig. 1 \quad Feynman diagrams describing the Compton 
excitation of a nucleus.}
\end{center}

\vspace{3in}

\begin{center}
\parbox{\textwidth}{Fig. 2 \quad Diagrams describing the ``Compton'' effect on 
the bound electron (IEB mechanism).}
\end{center}

\newpage

\vspace*{3in}

\begin{center}
\parbox{\textwidth}{Fig. 3 \quad Scheme of an Anti-Stokes $\beta$-transition on
a nucleus.}
\end{center}

\vspace{3in}

\begin{center}
\parbox{\textwidth}{Fig. 4 \quad Scheme of an Anti-Stokes electromagnetic 
transition on a nucleus.}
\end{center}

\newpage

\vspace*{4.0in}

\begin{center}
\parbox{\textwidth}{Fig. 5 \quad The quantity $T(\beta , {\Delta \over 
\hbar\omega_L})$ defined
by Eq. (25) (circular polarization) is defined by the relation $I = 8.73 
\cdot 10^{11}\beta_0^2$.  The curves a-d correspond to $\hbar\omega_L$ =
5 eV.  The nuclei $^{235}U$ and $^{183}W$ are denoted by their electronic
shell (see Table 2).  The curves c and g denoted by K correspond to the 
$^{105m}Ag$ (K shell) \cite{Kal88}.}
\end{center}

\vspace{2.0in}

\parbox{\textwidth}{Fig. 6 \quad Feynman diagrams corresponding to the
Electron Bridge mechanism.}
\newpage

\newpage

\vspace*{5.5in}

\parbox{\textwidth}{Fig. 7 \quad (a) The spectrum obtained after subtraction of
contributions from room background, source impurities, and external
bremsstrahlung.  The full circles represent experimental data while the full
line represents the total estimated spectrum $N(E_{\gamma})$.  The dashed line
represents the ICE contribution in the region of interest; at lower energies 
there are only ICE contributions. (b) The spectrum produced by the 
inelastic-electronic-bridge effect with 28.2 keV photons \cite{Kek85}.}

\newpage

\vspace*{3.0in}

\parbox{\textwidth}{Fig. 8 \quad Scheme explaining the elimination of 
the mismatch between the atomic and nuclear level energy differences by laser
radiation.}

\vspace{3.0in}

\parbox{\textwidth}{Fig. 9 \quad Feynman diagram describing a laser-assisted
resonance deexcitation of a nucleus.}

\newpage

\vspace*{3.0in}
\parbox{\textwidth}{Fig. 10 \quad $F_K$ vs. $I$. (a) Two photon case ($\hbar
\omega_L$ = 2.32 eV), (b) Four photon case ($\hbar\omega_L$ = 1.32 eV).
Curves in both figures are numbered by K \cite{Ka912}.}

\vspace{3.0in}

\parbox{\textwidth}{Fig. 11 \quad $f_K$ vs $K$. (a) and (b) indicate different
channel structure of $f_K$ at $I = 10^{11.5} {W \over cm^2}$ and $10^2 
{W \over cm^2}$ \cite{Ka912}.}

\newpage

\vspace*{1in}

\centerline{Fig. 12}

\vspace{3.0in}

\parbox{\textwidth}{(a) Scheme of laser-assisted resonance excitation of 
a nucleus.}

\vspace{2.0in}

\parbox{\textwidth}{(b)Feynman diagram corresponding to the scheme of Fig. 
14(a).}

\vspace{0.5in}

\centerline{Fig. 13}

\newpage

\vspace*{1.75in}

\parbox{\textwidth}{Fig. 14 \quad The effective Feynman diagram describing the 
laser-assisted resonant nuclear excitation.}

\vspace{2.5in}

\parbox{\textwidth}{Fig. 15 \quad A ``dressed'' (by laser radiation) 
intermediate electronic state.}

\vspace*{1.75in}

\parbox{\textwidth}{Fig. 16 \quad Scheme of the two lowest levels of the
thorium-229 nucleus.}

\newpage

\vspace*{5.5in}

\parbox{\textwidth}{Fig. 17 \quad Deexcitation probability for the 
${3 \over 2}^+$
state in the $^{229}Th$ nucleus for the $7s_{1 \over 2} \rightarrow 8s_{1 \over
2}$ electronic transitions as a function of laser intensity.  The photon
energy is fixed to the energy of a resonant two-photon process for a nuclear 
excitation energy of 3.5 eV (solid line), 4.0 eV (long-dashed line), and 3.0 eV
(short-dashed line) or the resonant four-photon process for an energy of 3.5 eV
(dotted line) \cite{Typ96}.}

\newpage

\vspace*{4.0in}

\parbox{\textwidth}{Fig. 18 \quad Scheme of the laser-assisted energy and
angular momentum transfer from the atomic shell to the $^{229}Th$ nucleus.}    
\end{document}